\documentclass[11pt]{article}
\usepackage{amssymb}
\usepackage{graphicx}
\usepackage[mathscr]{eucal}
\begin{document}
\newtheorem{prop}{}[section]
\newtheorem{defi}[prop]{}
\newtheorem{lemma}[prop]{}
\newtheorem{rema}[prop]{}
\newcommand{\boma}[1]{{\mbox{\boldmath $#1$} }}
\newcommand{\E}[1]{ {\mathcal E}^{#1} }
\def\mat{{\frak g}}
\def\la{\lambda}
\def\tG{t_{\scriptscriptstyle{G}}}
\def\tN{t_{\scriptscriptstyle{N}}}
\def\TK{t_{\scriptscriptstyle{K}}}
\def\CK{C_{\scriptscriptstyle{K}}}
\def\CN{C_{\scriptscriptstyle{N}}}
\def\CG{C_{\scriptscriptstyle{G}}}
\def\CCG{{\mathscr{C}}_{\scriptscriptstyle{G}}}
\def\tf{{\tt f}}
\def\ti{{\tt t}}
\def\ta{{\tt a}}
\def\tc{{\tt c}}
\def\tF{{\tt R}}
\def\P{{\mathscr P}}
\def\es{{\mathcal \SS}}
\def\TI{\tilde{I}}
\def\TJ{\tilde{J}}
\def\Lin{\mbox{Lin}}
\def\Hinfc{ H^{\infty}(\reali^d, \complessi) }
\def\Hnc{ H^{n}(\reali^d, \complessi) }
\def\Hmc{ H^{m}(\reali^d, \complessi) }
\def\Hac{ H^{a}(\reali^d, \complessi) }
\def\Dc{\DD(\reali^d, \complessi)}
\def\Dpc{\DD'(\reali^d, \complessi)}
\def\Sc{\SS(\reali^d, \complessi)}
\def\Spc{\SS'(\reali^d, \complessi)}
\def\Ldc{L^{2}(\reali^d, \complessi)}
\def\Lpc{L^{p}(\reali^d, \complessi)}
\def\Lqc{L^{q}(\reali^d, \complessi)}
\def\Lrc{L^{r}(\reali^d, \complessi)}
\def\Hinfr{ H^{\infty}(\reali^d, \reali) }
\def\Hnr{ H^{n}(\reali^d, \reali) }
\def\Hmr{ H^{m}(\reali^d, \reali) }
\def\Har{ H^{a}(\reali^d, \reali) }
\def\Dr{\DD(\reali^d, \reali)}
\def\Dpr{\DD'(\reali^d, \reali)}
\def\Sr{\SS(\reali^d, \reali)}
\def\Spr{\SS'(\reali^d, \reali)}
\def\Ldr{L^{2}(\reali^d, \reali)}
\def\Hinfk{ H^{\infty}(\reali^d, \KKK) }
\def\Hnk{ H^{n}(\reali^d, \KKK) }
\def\Hmk{ H^{m}(\reali^d, \KKK) }
\def\Hak{ H^{a}(\reali^d, \KKK) }
\def\Dk{\DD(\reali^d, \KKK)}
\def\Dpk{\DD'(\reali^d, \KKK)}
\def\Sk{\SS(\reali^d, \KKK)}
\def\Spk{\SS'(\reali^d, \KKK)}
\def\Ldk{L^{2}(\reali^d, \KKK)}
\def\Knb{K^{best}_n}
\def\sc{\cdot}
\def\k{\mbox{{\tt k}}}
\def\x{\mbox{{\tt x}}}
\def\g{ {\textbf g} }
\def\QQQ{ {\textbf Q} }
\def\AAA{ {\textbf A} }
\def\UUU{ {\textbf U} }
\def\gr{\mbox{graph}~}
\def\Q{\mbox{Q}_a}
\def\loc{\mbox{loc}}
\def\K{\Gamma}
\def\PZ{{\Lambda}}
\def\PZAL{\mbox{P}^{0}_\alpha}
\def\PL{{L}}
\def\PU{\mbox{P}^{1}_a}
\def\PD{\mbox{P}^{2}_a}
\def\PK{\mbox{P}^{k}_a}
\def\PKU{\mbox{P}^{k+1}_a}
\def\PI{\mbox{P}^{i}_a}
\def\Pell{\mbox{P}^{\ell}_a}
\def\PTM{\mbox{P}^{3/2}_a}
\def\AZ{A_{(r)}}
\def\AU{U}
\def\epsilona{\epsilon^{\scriptscriptstyle{<}}}
\def\epsilonb{\epsilon^{\scriptscriptstyle{>}}}
\def\lgraffa{ \mbox{\Large $\{$ } \hskip -0.2cm}
\def\rgraffa{ \mbox{\Large $\}$ } }
\def\restriction{\upharpoonright}
\def\M{{\scriptscriptstyle{M}}}
\def\m{m}
\def\Fre{Fr\'echet~}
\def\I{{\mathcal N}}
\def\ap{{\scriptscriptstyle{ap}}}
\def\fiap{\varphi_{\ap}}
\def\dfiap{{\dot \varphi}_{\ap}}
\def\DDD{ {\mathfrak D} }
\def\BBB{ {\textbf B} }
\def\EEE{ {\textbf E} }
\def\FFF{ {\textbf F} }
\def\GGG{ {\textbf G} }
\def\TTT{ {\textbf T} }
\def\KKK{ {\textbf K} }
\def\HHH{ {\textbf K} }
\def\Ee{ {\tt E} }
\def\Ff{ {\mathscr F} }
\def\Gg{ {\mathscr G} }
\def\FFi{ {\bf \Phi} }
\def\GGam{ {\bf \Gamma} }
\def\sc{ {\scriptstyle{\bullet} }}
\def\a{a}
\def\ep{\epsilon}
\def\c{\kappa}
\def\C{K}
\def\parn{\par\noindent}
\def\teta{M}
\def\elle{L}
\def\ro{\rho}
\def\al{\alpha}
\def\si{\sigma}
\def\be{\beta}
\def\ga{\gamma}
\def\de{\delta}
\def\te{\vartheta}
\def\ch{\chi}
\def\et{\eta}
\def\complessi{{\textbf C}}
\def\reali{{\textbf R}}
\def\interi{{\textbf Z}}
\def\naturali{{\textbf N}}
\def\T{{\textbf T}}
\def\T1{{\textbf T}^{1}}
\def\ee{{E}}
\def\EE{{\mathcal E}}
\def\FF{{\mathcal F}}
\def\GG{{\mathcal G}}
\def\KK{{\mathcal K}}
\def\PP{{\mathcal P}}
\def\QQ{{\mathcal Q}}
\def\J{J}
\def\Np{{\hat{N}}}
\def\Lp{{\hat{L}}}
\def\Jp{{\hat{J}}}
\def\Pp{{\hat{P}}}
\def\Pip{{\hat{\Pi}}}
\def\Vp{{\hat{V}}}
\def\Ep{{\hat{E}}}
\def\Fp{{\hat{F}}}
\def\Gp{{\hat{G}}}
\def\Kp{{\hat{K}}}
\def\Ip{{\hat{I}}}
\def\Tp{{\hat{T}}}
\def\Mp{{\hat{M}}}
\def\La{\Lambda}
\def\Ga{\Gamma}
\def\Si{\Sigma}
\def\Upsi{\Upsilon}
\def\Gam{\Gamma}
\def\Gag{{\check{\Gamma}}}
\def\Lap{{\hat{\Lambda}}}
\def\Sip{{\hat{\Sigma}}}
\def\Upsig{{\check{\Upsilon}}}
\def\Kg{{\check{K}}}
\def\ellp{{\hat{\ell}}}
\def\j{j}
\def\jp{{\hat{j}}}
\def\BB{{\mathcal B}}
\def\LL{{\mathcal L}}
\def\SS{{\mathcal S}}
\def\DD{{\mathcal D}}
\def\VV{{\mathcal V}}
\def\WW{{\mathcal W}}
\def\OO{{\mathcal O}}
\def\RR{{\mathcal R}}
\def\TT{{\mathcal T}}
\def\AA{{\mathcal A}}
\def\CC{{\mathcal C}}
\def\JJ{{\mathcal J}}
\def\NN{{\mathcal N}}
\def\WW{{\mathcal W}}
\def\HH{{\mathcal H}}
\def\XX{{\mathcal X}}
\def\YY{{\mathcal Y}}
\def\ZZ{{\mathcal Z}}
\def\UU{{\mathcal U}}
\def\CC{{\mathcal C}}
\def\XX{{\mathcal X}}
\def\RR{{\mathcal R}}
\def\cir{{\scriptscriptstyle \circ}}
\def\circa{\thickapprox}
\def\vain{\rightarrow}
\def\parn{\par \noindent}
\def\salto{\vskip 0.2truecm \noindent}
\def\spazio{\vskip 0.5truecm \noindent}
\def\vs1{\vskip 1cm \noindent}
\def\fine{\hfill $\diamond$ \vskip 0.2cm \noindent}
\newcommand{\rref}[1]{(\ref{#1})}
\def\beq{\begin{equation}}
\def\feq{\end{equation}}
\def\beqq{\begin{eqnarray}}
\def\feqq{\end{eqnarray}}
\def\barray{\begin{array}}
\def\farray{\end{array}}
\makeatletter \@addtoreset{equation}{section}
\renewcommand{\theequation}{\thesection.\arabic{equation}}
\makeatother
\begin{titlepage}
\begin{center}
{\huge On approximate solutions of semilinear evolution
equations.}
\end{center}
\vspace{0.5truecm}
\begin{center}
{\large
Carlo Morosi${}^1$, Livio Pizzocchero${}^2$} \\
\vspace{0.5truecm} ${}^1$ Dipartimento di Matematica, Politecnico
di
Milano, \\ P.za L. da Vinci 32, I-20133 Milano, Italy \\
e--mail: carmor@mate.polimi.it \\
${}^2$ Dipartimento di Matematica, Universit\`a di Milano\\
Via C. Saldini 50, I-20133 Milano, Italy\\
and Istituto Nazionale di Fisica Nucleare, Sezione di Milano, Italy \\
e--mail: livio.pizzocchero@mat.unimi.it
\end{center}
\begin{abstract}
A general framework is presented to discuss the approximate solutions
of an evolution equation in a Banach space, with a linear part generating a semigroup
and a sufficiently smooth nonlinear part. A theorem is presented, allowing
to infer from an approximate solution the existence of an exact solution. According to this theorem,
the interval of existence of the exact solution and the distance of the latter
from the approximate solution can be evaluated
solving a one-dimensional "control" integral equation, where the unknown
gives a bound on the previous distance as a function of time.
For example, the control equation can be
applied to the approximation methods based on the reduction of the evolution
equation to finite-dimensional manifolds: among them, the Galerkin method is
discussed in detail. To illustrate this framework, the nonlinear
heat equation is considered. In this case the control equation is used
to evaluate the error of the Galerkin approximation; depending on the
initial datum, this approach either grants global existence of the solution or
gives fairly accurate bounds on the blow up time.
\end{abstract}
\vspace{0.2cm} \noindent
\textbf{Keywords:} Differential equations, theoretical approximation, nonlinear heat equation,
blow up.
\par \vspace{0.05truecm} \noindent \textbf{AMS 2000 Subject classifications:} 34AXX, 35AXX, 35KXX. \par
\vspace{0.5truecm} \noindent
\textbf{To appear in "Reviews in Mathematical Physics".}
\end{titlepage}
\setcounter{footnote}{0}
\section{Introduction.}
In this paper we consider, within a Banach space $\FFF$,
a Volterra integral equation
\beq \varphi(t) = \UU(t - t_0)
f_0 + \int_{t_0}^t~ d s~ \UU(t - s) \PP(\varphi(s),s)~, \label{int0}
\feq
for an unknown function $\varphi$ from a real interval to $\FFF$.
Here $f_0 \in \FFF$, $\UU$ is a linear semigroup on
$\FFF$ and $\PP$ is a locally Lipschitz nonlinear map from an
open set of $\FFF \times \reali$ to $\FFF$. If $\UU$ is the semigroup generated by a linear
operator $\AA: Dom \AA \subset \FFF \vain \FFF$, under minimal
technical conditions the above Volterra equation is
equivalent to a Cauchy problem
\beq {\dot \varphi}(t) = \AA \varphi(t) + \PP(\varphi(t),t), \qquad
\varphi(t_0) = f_0~, \qquad (\dot{~} := d/ d t). \label{si1} \feq
To standardize the language, problems \rref{int0},
\rref{si1} are defined precisely in Sect.\ref{prelim};
local existence and uniqueness of their solutions are well known. \parn
The aim of this paper is to discuss the approximate solutions of
\rref{int0}. In the most general sense, an approximate solution is
simply a continuous map $t \mapsto \fiap(t)$ which can be inserted
in the r.h.s. of \rref{int0}, i.e., such that $\gr \fiap \subset Dom \PP$.
For any such map, we can define the \textsl{integral error}
as the difference between the two sides of \rref{int0}. If
$\fiap$ is a bit more regular, the integral error is determined by the
\textsl{differential} and \textsl{datum errors} which are, respectively, the
differences between the two sides in the differential
equation and in the initial condition of \rref{si1}.
\parn
All the above concepts are formalized in Sect.\ref{teoria}. Here, we also present a general statement
(Prop.\ref{main}) which
can be applied to an approximate solution $t \mapsto \fiap(t)$
to infer the existence of an \textsl{exact} solution $\varphi$
on an appropriate time interval, and also to estimate the difference
$\varphi(t) - \fiap(t)$. The essential character in Prop.\ref{main}
is an integral \textsl{control inequality}, depending on the available estimators for the
integral error of $\fiap$ and for the growth of $\PP$ away from the graph of $\fiap$. \parn
The unknown in the control inequality is a real,
nonnegative function $t \mapsto R(t)$; if
a solution $R$ is found to exist on a time interval $[t_0, t_1|$ (i.e., either
$[t_0, t_1]$ or $[t_0, t_1)$), then
it is granted that \rref{int0} possesses an exact
solution $\varphi : [t_0, t_1| \vain \FFF$, and that $\| \varphi(t) - \fiap(t) \| \leq R(t)$. \parn
In typical cases, a solution of the previous integral inequality
can be constructed solving an ordinary differential
equation for $R$, that we call as well the \textsl{control equation}.
In this way, the problem of giving estimates on the
existence time for \rref{int0} and on its exact solution
$\varphi$, living in $\FFF$ which is typically of infinite dimension,
is reduced to the analysis of a \textsl{one-dimensional} ODE.
\parn
Prop.\ref{main} can be regarded as a general formulation
of many statements about specific evolutionary problems,
often encountered in the literature. From this viewpoint, the
content of this Proposition is not at all surprising: however, the
technique we use to prove it is essentially
different from the arguments often employed in related situations.
The standard way of thinking would suggest to prove Prop \ref{main}
in two steps: a) derive (via some nonlinear Gronwall lemma
\cite{Wal}) an a priori bound $\| \varphi(t) - \fiap(t) \| \leq R(t)$,
holding until $\varphi(t)$ exists; b) show that nonexistence of $\varphi$
on the whole interval $[t_0, t_1|$ would contradict the previous bound:
this argument is called the "continuation principle" in \cite{Zei2}. \parn
On the contrary,
the proof we propose (in Sect.\ref{prova}) is
very direct, and shows that $\varphi$ can be constructed on the
whole $[t_0, t_1|$ by a convergent Peano-Picard iteration, applying
repeatedly the Volterra integral operator to the approximate
solution $\fiap$. The control inequality ensures the invariance under
the Volterra operator of the space of functions with distance $\leq R(t)$ from $\fiap(t)$
on $[t_0, t_1|$; the confinement to this domain of all iterates of $\fiap$,
and the local Lispchitz nature of $\PP$, allow to
prove their convergence to a function $\varphi$, also
distant less than $R$ from $\fiap$.
\parn
As a first, very simple illustration of Prop.\ref{main}, in Sect.\ref{elem} we
apply the control equation to the approximate solution $\fiap(t) :=0$.
In spite of the trivial choice for $\fiap$, the control equation
gives useful information on the interval of existence
and on the growth of the exact solution $\varphi$, depending
on the norm $\| f_0 \|$ of the initial datum. The accuracy of these predictions is tested on an example, concerning the
(one-dimensional) wave equation with polynomial nonlinearity. \parn
A second, more refined application of the control equation is proposed
in Sect.\ref{galerk} for the Galerkin scheme (and similar approaches). In the conventional
formulation, the Galerkin method is an algorithm to construct approximate
solutions $t \mapsto \fiap(t)$ of \rref{int0} with values in a finite-dimensional submanifold
of $\FFF$. In this Section, the standard evolution equations for the coordinates of $\fiap(t)$
in the Galerkin submanifold are coupled with the
control equation for $R(t)$; in this way,
a finite-dimensional system of ODE's gives simultaneously the Galerkin
approximate solution $\fiap$, an interval $[t_0, t_1|$ on which the exact
solution $\varphi$ of \rref{int0} is granted to exist and an upper bound
for $\| \fiap(t) - \varphi(t) \|$ on this interval. \parn
In Sect.\ref{appl}, all the previous results are applied
to a nonlinear heat equation, working for
simplicity in one space dimension (with
a spatial coordinate $x \in (0,\pi)$). In this case, the
Cauchy problem \rref{si1}  (with initial time $t_0 :=0$) is, symbolically,
\beq {\dot \varphi}(x,t) = \varphi_{x x}(x,t) + \varphi(x,t)^p~, \qquad \varphi(x,0) = f_0(x) \label{symb} \feq
with $p \in \{2,3,4,...\}$, to be discussed in the Sobolev space $\FFF := H^1_0(0,\pi)$.
The implementation of the general framework in the present case with polynomial
nonlinearity requires accurate information on the pointwise product of functions in
$H^{1}_{0}(0,\pi)$; in particular, precise estimates are needed for the norm
$\| f h \|$ when $f, h$ are in this space (see the Appendix \ref{appe1} about this,
and \cite{MP} for more general information about
multiplication in Sobolev spaces). \parn
To exemplify some general facts about \rref{symb}, in the same Section we consider the
initial datum $f_0(x) := \sqrt{2/\pi} A \sin x$. If the (nonnegative) constant $A$ is below a critical value,
the control equation for the zero approximate solution suffices to
prove existence of a globally defined solution $\varphi : [0,+\infty) \vain \FFF$ of
\rref{symb}. For larger $A$,
the same control equation gives a finite lower bound for the existence time
of the solution $\varphi$. These conclusions are
complementary to the ones arising from a known "blow up" theorem of Kaplan
for the nonlinear heat equation (see \cite{Kap}; a review is given in the
Appendix \ref{appe2}). When Kaplan' s theorem is applied to \rref{symb}
with the previous datum, for sufficiently large $A$ it predicts a finite, explicitly
determined upper bound on the existence time of the solution. \parn
Again in Sect.\ref{appl}, we add to the above facts the information arising from application of the control
equation to the Galerkin scheme; the chosen Galerkin submanifold is the linear span of finitely many elements in the
Fourier basis. As an example, we consider the Galerkin differential equations for two modes,
coupled with the control equation for $R$, with $p=2$ and the previous $f_0$.
This system in three unknown real functions can be
easily treated by any package for the numerical solution of ODE's; the results obtained
by the MATHEMATICA package, for several values of $A$, are presented with some detail.
Among other things, the Galerkin approach with the control equation allows to increase the critical
value of $A$ below which global existence is granted for \rref{symb}; for
$A$ above the new critical value, a better lower bound for the existence time is derived.
If $A$ is fairly large, the new lower bound is close to the Kaplan upper bound,
which yields an uncertainty between $20 \%$ and $30 \%$ on the existence time
of the exact solution. Also, the upper bound $R(t)$ on $\| \varphi(t) - \fiap(t) \|$
is fairly small in comparison with $\| \fiap(t) \|$ for non large $t$.
\parn
To some extent, it is surprising that a fairly good accuracy can be obtained combining
the control equation with a Galerkin scheme in two modes only. These outcomes
encourage us to hope that the same method would give nontrivial information
on the Cauchy problem for the equations of fluid dynamics, whose Galerkin approximations
in few modes give rise, among others, to the widely studied Lorentz model \cite{Lor} \cite{Tem}.
\section{Preliminaries.}
\label{prelim}
Throughout the paper, $\FFF$ denotes a real or complex
Banach space with norm $\|~\|$ and elements $f, f_0, f_1, h, ...$~.
We write $\BBB(f_0, \rho)$ for the open ball in $\FFF$ of center $f_0$ and radius
$\rho$ (if $\rho = +\infty$, this means the whole $\FFF$).
Let us be given  a linear operator
\beq \AA : Dom \AA \subset \FFF \vain \FFF \feq
with domain a linear subspace of $\FFF$; whenever we speak of a
continuous map from/to $Dom \AA$, we always refer to the topology
of the graph norm $\| f \|_{\AA} := \| f \| + \| A f \|$
(as well known, $Dom \AA$ is complete in this norm if and only
if $\AA$ is closed).
We denote with $\LL(\FFF)$ the Banach space of bounded linear operators
of (the whole) $\FFF$ into itself. \parn
We always write $[t_0, t_1|$ for a real interval of the form
$[t_0, t_1]$ or $[t_0, t_1)$ (always intending $t_0 < t_1$; in the second case, $t_1$ can be $+\infty$).
If $\psi : [t_0, t_1| \vain \FFF$,
the graph of this function and the tube around $\psi$ of any radius $\rho \in (0,+\infty]$ are
\beq \gr \psi := \{~ (\psi(t), t)~|~t \in [t_0, t_1|~\} \subset \FFF \times \reali~. \feq
\beq \TTT(\psi, \rho) := \{ (f, t) \in \FFF \times [t_0, t_1|~~|~~
\| f - \psi(t) \| < \rho \} \label{detub} \feq
(the latter is the whole $\FFF \times [t_0, t_1|$, if $\rho = + \infty$; it becomes
$\BBB(f_0, \rho) \times [t_0, t_1|$, if $\psi(t)=$ const. $=f_0$).
\vskip 0.2cm \noindent
\textbf{Linear semigroups on $\boma{\FFF}$.} This name indicates maps $\UU$ such that
\beq \UU : [0, +\infty) \vain \LL(\FFF)~, \qquad
\UU(t+s) = \UU(t) \, \UU(s),~~ \UU(0) = \bf{1}_{\FFF}~. \feq
The \textsl{generator} of a linear semigroup $\UU$ is the linear operator
\beq A : Dom \AA \subset \FFF \vain \FFF~, \qquad f \mapsto \AA f~, \feq
\beq Dom \AA := \{ f \in \FFF~|~\left. {d \over d t} \right|_{t=0}
\Big[ \UU(t) f \Big]~\mbox{exists} \}~, \qquad
\AA f := \left. {d \over d t} \right|_{t=0}
\Big[ \UU(t) f \Big] \feq
(with $(d / d t)_{t=0}$ denoting the right derivative). \parn
A linear semigroup $\UU$ in $\FFF$ is strongly continuous if for all $f \in \FFF$
the map $[0, +\infty) \vain \FFF$, $t \mapsto \UU(t) f$ is
continuous. In this case (see, e.g., \cite{Caz} \cite{Zei2}),
the map $(f, t) \mapsto \UU(t) f$ is jointly continuous,
the generator $\AA$ is densely defined in $\FFF$ and closed, and a) b) hold: \parn
a) $\AA$ determines $\UU$. For all $f_0 \in Dom \AA$,
the function $t \mapsto \UU(t) f_0$ is the unique function $\varphi$ such that
\beq \varphi \in C([0, +\infty), Dom \AA) \cap C^1([0,+\infty), \FFF),~~
\dot{\varphi}(t) = \AA \varphi(t)~\mbox{for all $t$},~~\varphi(0) =
f_0~; \label{iden0} \feq
b) for any function  $\psi \in C([t_0, t_1|, Dom \AA) \cap C^1([t_0, t_1|, \FFF)$
and $t$ in this interval, it is
\beq \psi(t) = \UU(t - t_0) \psi(t_0) + \int_{t_0}^t d s~
\UU(t - s) \left[ {\dot \psi}(s) - \AA \psi(s) \right]
\label{iden} \feq
(here and in the sequel, the dot indicates the derivative).
A linear semigroup $\UU$ is uniformly continuous if the map
$\UU$ is continuous from $[0,+\infty)$ to $\LL(\FFF)$ with the standard operator norm
(uniformly continuous semigroup); this happens if and only if
$\UU$ has generator $\AA \in \LL(\FF)$, and gives
a trivial example of strongly continuous semigroup (extendable to $t < 0$).
\begin{prop}
\label{nob} \textbf{Definition.} An \textsl{estimator} for a
linear semigroup $\UU$ on $\FFF$ is a continuous
function $u :[0,+\infty) \vain [0, +\infty)$ such that, for all $f \in \FFF$ and $t \in [0,+\infty)$,
\beq \qquad \qquad \qquad \qquad \qquad
\qquad \| \UU(t) f \| \leq u(t) \| f \|~. \qquad \qquad \qquad \qquad \qquad \qquad \qquad \diamond \label{a1} \feq
\end{prop}
Each strongly continuous linear semigroup admits an estimator of the form $u(t) = U
e^{-B t}$, where $U \geq 1$ and $B$ are real constants: see
\cite{Dun}.
\vskip 0.2cm \noindent
\textbf{Lipschitz maps.} If $\CC$, $\DD$ are subsets of a topological vector space,
we say that $\CC$ is a \textsl{strict subset of} $\DD$, and write
$\CC \Subset \DD$, if $\CC$ is bounded and $\overline{\CC} \subset \DD$,
the symbol $\overline{{~}^{~}}$ denoting the closure.
Now, let us be given a (possibly nonlinear) map, with open domain,
\beq \PP : Dom \PP \subset \FFF \times \reali \vain \FFF~, \qquad (f, t) \mapsto \PP(f, t)~.
\label{pp} \feq
\begin{prop}
\label{lip}
\textbf{Definition.} We say that $\PP$ is \textsl{Lipschitz at fixed time} (or, respectively, \textsl{Lipschitz})
on the \textsl{strict subsets} of its domain if, for every $\CC \Subset Dom \PP$, there is a nonnegative
constant $L=L(\CC)$ (or, resp., a pair of nonnegative constants $L = L(\CC)$, $M = M(\CC)$) such that
\beq \| \PP(f, t) - \PP(f', t) \| \leq L \| f - f' \| \qquad \mbox{for $(f, t), (f', t) \in \CC$;}
\label{llipf} \feq
\beq \| \PP(f, t) - \PP(f', t') \| \leq L \| f - f' \| + M | t - t' |
\qquad \mbox{for $(f, t), (f', t') \in \CC$~.} \hspace{1cm} \diamond
\label{llippff} \feq
\end{prop}
Of course \rref{llippff} implies \rref{llipf} and the continuity of $\PP$.
\vskip 0.2cm \noindent
\textbf{An example.} \label{exa}
Some applications presented in the sequel rely on a map $\PP$ of the form
\beq Dom \PP = \FFF \times \Delta \qquad \mbox{($\Delta \subset \reali$ an open interval)}, ~~
\PP(f, t) := \P(f,...,f, t)~, \feq
where
\beq \P : \times^p \FFF \times \Delta \vain \FFF~~(p \in \{1,2,...\})~,
\qquad (f_1,...,f_p\,,t) \vain \P(f_1,...,f_p\,,t) \feq
is $\reali$-linear in each argument $f_1,...,f_p$\,; it is also assumed that
\beq \| \P(f_1, ..., f_p\,,t) \| \leq P(t)~ \| f_1 \| ... \| f_p \|~, \label{assu} \feq
$$ \| \P(f_1, ..., f_p\,,t) - \P(f_1, ..., f_p\,,t') \| \leq Q(t, t')~ \| f_1 \| ... \| f_p \| $$
for all $f_1, ..., f_p \in \FFF$ and $t, t' \in \Delta$, where $P : \Delta \vain [0,+\infty)$ and
$Q : \Delta \times \Delta \vain [0,+\infty)$ are continuous functions.
It is finally required that, for each $B \Subset \Delta$, there is a
constant $M = M(B)$ such that
\beq \hspace{4cm} Q(t, t') \leq M | t - t' | \qquad \mbox{for $t, t' \in B$}. \hspace{3cm} \diamond \label{qtt} \feq
\begin{prop}
\label{inep}
\textbf{Proposition.} With the previous assumptions,
for all $f, f' \in \FFF$ and $t, t' \in \Delta$ it is
\beq \| \PP(f, t) - \PP(f', t') \| \leq P(t) \sum_{j=1}^{p} \left( \barray{cc} p \\ j \farray \right)
\| f' \|^{p-j} \| f - f' \|^j + Q(t, t') \| f' \|^p~. \label{zero} \feq
\end{prop}
\textbf{Proof.} Setting for convenience $f_1 := f'$, $f_2 := f - f'$ we can write
\beq \PP(f, t) - \PP(f', t) = \P(f_1 + f_2, ..., f_1 + f_2, t) -
\P(f_1, ..., f_1, t) = \feq
$$ = \sum_{j=1}^p \! \! \sum_{~(l_1, ..., l_p) \in \Lambda_{p j}} \P(f_{l_1}, ..., f_{l_p}, t), \quad
\Lambda_{p j} := \{ (l_1,...,l_p) \in \{1,2\}^p~|~\mbox{$l_s = 2$ for $j$ values of $s$}~\}~. $$
From the first inequality \rref{assu}, we infer
\beq \| \PP(f, t) - \PP(f', t) \| \leq
P(t) \sum_{j=1}^p \left( \barray{cc} p \\ j \farray \right) \| f' \|^{p-j} \| f - f' \|^j~,
\label{tre}
\feq
because $\Lambda_{p j}$ has cardinality $\scriptscriptstyle{\left( \barray{cc} p \\ j \farray \right)}$.
Finally, the second assumption \rref{assu} gives
\beq \| \PP(f', t) - \PP(f', t') \| \leq Q(t, t') \| f' \|^p \label{due} \feq
and \rref{tre} \rref{due}, with the triangular inequality, yield the thesis \rref{zero}.
\fine
Eq.\rref{zero} will be frequently used in the sequel; together with \rref{qtt}, it implies
\begin{prop}
\textbf{Corollary.} The map $\PP$ is Lipschitz on the strict subsets of $\FFF \times \Delta$.
\fine
\end{prop}
\textbf{General formulation of the Volterra and Cauchy problems.}
We define formally both problems, and review their relations. \parn
\begin{prop}
\label{volter}
\textbf{Definition.} Let us be given: \parn i) a
strongly continuous linear semigroup $\UU$ on $\FFF$; \parn ii) a
continuous map $\PP : Dom \PP \subset \FFF \times \reali \vain \FFF$, with open domain; \parn
iii) a pair $(f_0, t_0) \in Dom \PP$.
\parn The \textsl{Volterra problem} related to $\UU$ and $\PP$ with
\textsl{datum $f_0$ at time $t_0$} is the following one:
$$ \mbox{\textsl{Find}}~
\varphi \in C([t_0, t_1|, \FFF) \quad \mbox{\textsl{such that $\gr
\varphi \subset Dom \PP$ and }}$$
\beq \varphi(t) = \UU(t - t_0)
f_0 + \int_{t_0}^t~ d s~ \UU(t - s) \PP(\varphi(s),s) \quad
\mbox{\textsl{for all} $t \in [ t_0, t_1|$}~. \label{int} \feq
\end{prop}
\begin{prop}
\label{decau}
\textbf{Definition.} Consider: \parn i) a linear
operator $\AA : Dom \AA \subset \FFF \vain \FFF$; \parn ii) a
continuous map $\PP$ as in the previous definition; \parn iii) a
pair $(f_0, t_0) \in Dom \PP$ such that $f_0 \in Dom \AA$. \parn
The \textsl{Cauchy problem} corresponding to $\AA, \PP$ with datum
$f_0$ at time $t_0$ is the following one:
$$ \mbox{\textsl{Find}}~
\varphi \in C([t_0, t_1|, Dom \AA) \cap C^1([t_0, t_1|, \FFF) ~\mbox{\textsl{such that $\gr
\varphi \subset Dom \PP$ and}} $$
\beq {\dot \varphi(t)} = \AA \varphi(t) + \PP(\varphi(t),t)
\quad \mbox{\textsl{for all} $t \in [t_0, t_1|$}~, \qquad
\varphi(t_0) = f_0 \label{cau} ~.\feq
\end{prop}
\begin{prop}
\label{integral} \textbf{Proposition.} Let $\AA, \PP, f_0, t_0$ be
as in Def.\ref{decau}, and further assume $\AA$ to be the
generator of a strongly continuous linear semigroup $\UU$. Then:
\parn
i) a solution $\varphi$ of the Cauchy problem
\rref{cau} is also solution of the Volterra problem
\rref{int}; \parn
ii) as a partial converse, a solution $\varphi$ of
the Volterra problem \rref{int} is a solution of the Cauchy problem
\rref{cau} in either of these situations: $\alpha$) $\FFF$ is reflexive and $\PP$ is
Lispchitz on the strict subsets of its domain; $\beta$) (trivial case) $\AA \in \LL(\FFF)$,
no further assumptions on $\FFF$ and $\PP$.
\end{prop}
\textbf{Proof.} It is essentially based on
(\ref{iden0}-\ref{iden}): see \cite{Caz} (conditions $\alpha$) $\beta$) in item  ii)
ensure a solution $\varphi \in C([t_0, t_1|,\FFF)$
of \rref{int} to be in $C^1([t_0, t_1|, \FFF) \cap C([t_0, t_1|, Dom \AA)$).
\fine
In particular, the operator $\AA := 0$ is the generator of the
identity semigroup $\UU(t) = \bf{1}_{\FFF}$ for all $t$.
With this remark, the framework of this paper applies to any
ODE ${\dot \varphi(t)} = \PP(\varphi(t),t)$ in a Banach space, also including
the finite dimensional cases $\FFF = \reali^m$ or $\complessi^m$.
\parn
\begin{prop}
\label{unique}
\textbf{Proposition.} Consider the Volterra problem \rref{int}, where $\UU$ is a
strongly continuous linear semigroup,
and $\PP$ is continuous and Lipschitz at fixed time on
the strict subsets of its domain; then i) ii) hold. \parn
i) Problem \rref{int} has a solution. \parn
ii) If $\varphi : [t_0, t_1|  \vain \FFF$ and $\varphi' :
[t_0, t_1'|  \vain \FFF$ are two solutions, it is
\beq \varphi(t) = \varphi'(t) \qquad \mbox{for $t \in [t_0, t_1| \cap [t_0,  t_1'|$}~. \feq
\end{prop}
\textbf{Proof.} ii) We consider any $t_2$ in the intersection of the domains.
Subtracting Eq.\rref{int} for $\varphi$ from the analogous
equation for $\varphi'$, and taking the norm, we obtain
\beq \| \varphi(t) - \varphi'(t) \| \leq \int_{t_0}^t d s~u(t - s)
\| \PP(\varphi(s), s) - \PP(\varphi'(s), s) \| \leq U L \int_{t_0}^t~\| \varphi(s) - \varphi'(s) \|~
\label{dfnt} \feq
for each $t \in [t_0, t_2]$. Here: $u$ is any estimator for $\UU$;
$U := \max_{s \in [t_0, t_2]} u(s)$; $L$ is a constant fulfilling the Lipschitz condition
\rref{llipf} for $\PP$ on the set $\CC := \gr (\varphi' \restriction [t_0, t_2])
\cup \gr(\varphi \restriction [t_0, t_2])$ (this $\CC$ is a strict subset of $Dom \PP$). Eq.\rref{dfnt} and
the classical Gronwall Lemma \cite{MPF} imply $\| \varphi(t) - \varphi'(t) \|=0$
for all $t \in [t_0, t_2]$. \parn
i) Eq.\rref{int} is the fixed point problem for a Volterra type integral operator, and a
solution can be constructed by standard Peano-Picard iteration, starting from
the function $\varphi_0(t) := \mbox{const.} := f_0$; see, e.g., \cite{Caz}. \fine
From our viewpoint, the previously mentioned argument for local existence is a particular case of a more
general statement, allowing to construct a solution of
\rref{int} by a Peano-Picard iteration with starting point any approximate solution
(of sufficiently small error); all this will be discussed in the next Section. \parn
Of course, Prop.\ref{integral} allows to transfer the statements on uniqueness
and existence from \rref{int} to \rref{cau}.
Let us call \textsl{maximal} a solution $\varphi$ of \rref{int} or \rref{cau}
which has no proper extension. If one can grant the existence of a solution
on a sufficiently small interval, and the coincidence of two solutions on the
intersection of their domains, it follows that a unique maximal solution $\varphi$ exists
and any other solution is a proper restriction of the maximal one. Furthermore, if local existence
is granted for \textsl{arbitray} data, the domain of the maximal solution $\varphi$
with a given datum has the form $[t_0, \vartheta)$ (otherwise,
$\varphi$ could be extended taking its value at $\vartheta$ as a new initial datum).
\vskip 0.2cm \noindent
\section{Approximate solutions. Statements of the main results.}
\label{teoria}
We consider a strongly continuous linear semigroup $\UU$ on the Banach space $\FFF$, and
a continuous function $\PP : Dom \PP \subset \FFF \times \reali \vain \FFF$ with
open domain. We are interested in the Volterra problem \rref{int}, for a given
pair $(f_0, t_0) \in Dom \PP$.
\begin{prop}
\label{allap} \textbf{Definition.} By an \textsl{approximate
solution} of problem \rref{int}, we mean any continuous function $\fiap
: [t_0, t_1| \vain \FFF$, such that $\gr \fiap \subset
Dom \PP$. Given any such function, we stipulate the following:
\parn
i) the \textsl{integral error} of $\fiap$ is the function
\beq E(\fiap) : t \in [t_0, t_1| \mapsto E(\fiap)(t) := \fiap(t) - \UU(t - t_0) f_0 -
\int_{t_0}^t d s~\UU(t - s) \PP(\fiap(s), s)~; \label{ier} \feq
an \textsl{integral error estimator} for $\fiap$ is a continuous
function $\EE : [t_0, t_1| \vain [0, +\infty)$ such that, for all $t$ in this interval,
\beq \| E(\fiap)(t) \| \leq \EE(t)~. \feq
ii) The \textsl{datum error} for $\fiap$ is the difference
\beq d(\fiap) := \fiap(t_0) - f_0~; \feq
a \textsl{datum error estimator} for $\fiap$ is a nonnegative real
number $\delta$ such that
\beq \| d(\fiap) \| \leq \delta~. \feq
iii) If $\AA$ is the generator of $\UU$ and
$\fiap \in C([t_0,t_1|, Dom \AA) \cap C^1([t_0, t_1|, \FFF)$,
the \textsl{differential error} of $\fiap$ is the function
\beq e(\fiap) : t \in [t_0, t_1| \mapsto e(\fiap)(t) := \dfiap(t) -
\AA \fiap(t) - \PP(\fiap(t), t)~; \feq
a \textsl{differential error estimator} for $\fiap$ is a continuous function $\ep : [t_0, t_1| \vain [0,+\infty)$
such that, for $t$ in this interval,
\beq \hspace{6cm} \| e(\fiap)(t) \| \leq \ep(t)~. \hspace{4cm} \diamond \feq
\end{prop}
Of course,
$\fiap$ is a solution of the Volterra (resp., Cauchy) problem iff $E(\fiap) = 0$
(resp., $d(\fiap) =0$ and $e(\fiap)=0$).
\begin{prop}
\label{lap} \textbf{Lemma.} Let $\fiap : [t_0, t_1| \vain \FFF$
be an approximate solution of the Volterra problem \rref{int}, and
assume the regularity conditions in item iii) of the previous
Definition. Then, the integral error of $\fiap$ is related to the
datum and differential errors by
\beq E(\fiap)(t) = \UU(t - t_0)\, d(\fiap) + \int_{t_0}^t d s~
\UU(t - s) e(\fiap)(s)~. \label{Ede}
\feq
If $\UU$, $d(\fiap)$, $e(\fiap)$
have estimators $u, \delta, \ep$, then $E(\fiap)$ has the estimator
\beq \EE(t) := u(t - t_0) \,\delta + \int_{t_0}^t d s~u(t - s) \ep(s) \qquad \mbox{for all
$t \in [t_0, t_1|$.} \label{hasthe} \feq
\end{prop}
\textbf{Proof.} Eq.\rref{Ede} follows applying the definitions of
$E(\fiap)$, $d(\fiap)$, $e(\fiap)$ and the identity \rref{iden} with $\psi := \fiap$.
Given \rref{Ede}, the bound \rref{hasthe} on $\| E(\fiap)(t) \|$ is evident. \fine
\textbf{Remark.} The estimator $\EE$
defined by \rref{hasthe} is useful, because in many cases it can be
easily computed. However, in peculiar situations involving
oscillating functions, this estimator can be rough. For
example, consider the semigroup $\UU(t) := \bf{1}_{\FFF}$ for all
$t$, with generator $\AA = 0$ and estimator $u(t):= 1$. Let us choose
$\fiap(t) := f_0$ for all $t$, so that
$d(\fiap) = 0$, $e(\fiap)(t) = - \PP(f_0, t)$ and $E(\fiap)(t) = -
\int_{t_0}^t d s~\PP(f_0, s)$. Suppose $\PP(f_0, t) = g_0 e^{i
\omega t}$, with $\omega \in (0, +\infty)$ and $g_0$ a vector of
the (complex) space $\FFF$; then, the best estimators for $d(\fiap)$ and $e(\fiap)$ are,
respectively, $\delta =0$ and $\ep(t) = \| g_0 \|$. Correspondingly,
Eq.\rref{hasthe} gives the integral error estimator
$\EE(t) = \| g_0 \| (t - t_0)$; on the other hand, it is
found by direct computation that $E(\fiap)(t) = i g_0 (e^{i
\omega t} - e^{i \omega t_0})/\omega$; thus $\| E(\fiap)(t) \|$
is a bounded function of $t$, whereas the estimator $\EE$ grows
linearly.
\parn
Similar drawbacks of the estimator \rref{hasthe} in the presence of oscillatory functions are met
(even for $\FFF = \reali^m$) if one considers a differential equation with fast periodic variables
and the approximate solutions which arise from averaging methods
\cite{Los}. \fine
To formulate the main theorem on approximate solutions, we need one more notion
describing the growth of $\PP$ away from a function
$\psi : [t_0, t_1| \vain \FFF$, such that $\gr \psi \subset Dom \PP$.
\begin{prop}
\label{grow}
\textbf{Definition.} A \textsl{growth estimator
for $\PP$ from $\psi$} (if it exists) is a continuous function
\beq \ell : [0, \rho) \times [t_0, t_1| \vain [0, +\infty)~, \qquad (r, t) \mapsto \ell(r, t)
\label{cont} \feq such that: \parn
i) $\rho \in (0, +\infty]$ and $\TTT(\psi, \rho) \subset Dom \PP$
(see Eq.\rref{detub}); \parn
ii) $\ell$ is nondecreasing in the first variable:
$\ell(r, t) \leq \ell(r', t)$ for $r \leq r'$; \parn
iii) for all $(f, t) \in \TTT(\psi, \rho)$, it is
\beq \hspace{3.5cm}
\| \PP(f, t) - \PP(\psi(t), t) \| \leq \ell(\| f - \psi(t) \|, t)~. \hspace{3cm} \diamond \label{esti} \feq
\end{prop}
\textbf{Remarks.} a) From a
continuous function $\ell' : [0, \rho) \times [t_0, t_1| \vain
[0,+\infty)$ fulfilling i) iii) but not ii), we can construct the function $\ell(r, t) :=$ $\max_{r' \in
[0, r]} \ell'(r', t)$, which also fulfils ii). \parn
b) A function $\PP$ which is Lipschitz at fixed time on a tube around $\psi$ possesses on it
a growth estimator linear in $r$. Less trivial estimators appear if $Dom \PP = \FFF \times \Delta$,
$\Delta$ a real interval, and one wishes to estimate the growth of $\PP$ from $\psi$ on the whole product space
$\FFF \times [t_0, t_1|$ (= on a tube of infinite radius). For instance, consider a map
$\PP$ as in the Example of page \pageref{exa}; the growth of $\PP$ from any
$\psi : [t_0, t_1| \subset \Delta \vain \FFF$
admits the estimator
\beq \ell(r, t) := P(t) \sum_{j=1}^p \left( \barray{cc} p \\ j \farray \right) \| \psi(t) \|^{p-j}~
r^j \feq
($r \in [0,+\infty)$, $t \in [t_0, t_1|$). To find this, apply Eq.\rref{zero}
with $f' = \psi(t)$ and $t' = t$.
\fine
We come to the main theorem of this Section: the proof will be given in Sect.\ref{prova}.
\begin{prop}
\label{main} \textbf{Proposition.} Let us be given a Volterra problem
\rref{int}, where: $\UU$ is a strongly continuous linear semigroup; $\PP : Dom \PP \subset
\FFF \times \reali \vain \reali$ is continuous and Lipschitz at fixed
time on the strict subsets of its open domain (Def. \ref{lip}). Assume that: \parn
i) $u$ is an estimator for $\UU$; \parn
ii) $\fiap : [t_0, t_1| \vain \FFF$ is an approximate solution of \rref{int},
and $\EE : [t_0, t_1| \vain [0,+\infty)$ is an estimator for the integral error $E(\fiap)$;
\parn
iii) $\ell : [0, \rho) \times [t_0, t_1| \vain [0, +\infty)$ is a growth estimator for
$\PP$ from $\fiap$ ($\rho \in (0, +\infty]$). \parn
Consider the following problem:
$$ \mbox{\textsl{Find}}~
R \in C([t_0, t_1|, [0, \rho)) \qquad \mbox{\textsl{such that}} $$
\beq \EE(t) + \int_{t_0}^t~ d s~ u(t - s)~ \ell(R(s), s) \leq R(t)
\quad \mbox{\textsl{for} $t \in [ t_0, t_1|$}~. \label{monod} \feq
If \rref{monod} has a solution $R$ with domain $[t_0, t_1|$, then \rref{int} has a
solution $\varphi$ with the same domain, and for all $t$ therein it is
\beq \| \varphi(t) - \fiap(t) \| \leq R(t)~. \feq
The solution $\varphi$ can be constructed
by a Peano-Picard iteration, starting from $\fiap$.
\fine
\end{prop}
\begin{prop}
\textbf{Definition.} Eq.\rref{monod} will be referred to as
the \textsl{control inequality}.
\fine
\end{prop}
\parn
\textbf{Remarks.} i) The function $R$ is required to exist on the same domain $[t_0, t_1|$
of $\fiap$. In many applications, one starts with an approximate solution on a domain
$[t_0, t_2|$ and then finds \rref{monod} to a have a solution $R$ on a domain
$[t_0, t_1| \subset [t_0, t_2|$; of course, in this case the previous Proposition must
be applied to $\fiap \restriction [t_0, t_1|$. \parn
ii) As anticipated, the argument we will employ to prove Prop.\ref{main} is different from the
"continuation principle" mentioned in the Introduction. Instead
of using a Gronwall Lemma plus a reductio ad absurdum, we will prove the existence
of $\varphi$ on $[t_0, t_1|$, and the bound \rref{monod}, by a constructive Peano-Picard
iteration; the convergence of this iteration on the whole $[t_0, t_1|$ has  some theoretical interest by itself.
Furthermore, this approach overcomes some technicalities required by the application of nonlinear Gronwall lemmas
(the analysis of the associated integral equation, and the necessity to determine
the greatest solution when uniqueness fails \cite{Wal}). \parn
Apart from the general concept of approximate solution employed here, the idea to prove
existence for an ODE $\dot f = \PP(f, t)$ by the Peano-Picard method, under
conditions of nonlinear growth for $\PP$ of more global type than the Lipschitz property
can be ascribed to Caratheodory \cite{Car}, and was developed in \cite{Waz} \cite{Ole}. \parn
iii) Of course, we can accept as a solution of \rref{monod} an
$R$ fulfilling the equation
\beq \EE(t) + \int_{t_0}^t~ d s~ u(t - s) \ell(R(s), s) = R(t)
\quad \mbox{\textsl{for} $t \in [ t_0, t_1|$}~,
\label{monodin} \feq
hereafter referred to as the \textsl{control integral equation}.
(The existence of such an $R$ on a sufficiently short
interval is granted by standard compactness arguments, see
\cite{Wal}. Uniqueness can be proved under supplementary assumptions of Lipschitz kind for
$\ell$).
\fine
Let us exploit a typical case, where the control integral equation \rref{monodin} is equivalent to a
Cauchy problem. To this purpose, assume that
\beq u(t) = U e^{-B t} \quad (U \geq 1, B \in \reali), \qquad
\EE(t) = U e^{-B(t - t_0)} \delta + U \int_{t_0}^t d s~ e^{-B(t - s)} \ep(s) \label{ee} \feq
for some constant $\delta \geq 0$ and some continuous function $\ep : [t_0, t_1| \vain [0,+\infty)$
(for example, the estimator $\EE$ derived from Eq.\rref{hasthe} has the above form).
Then, multiplying by $e^{B (t - t_0)}$ we see that
Eq.\rref{monodin} is equivalent to
\beq U \delta  + U \int_{t_0}^t d s~ e^{B(s - t_0)} \ep(s) + U
\int_{t_0}^t~ d s~e^{B(s - t_0)} \ell(R(s), s) = e^{B (t - t_0)} R(t)~. \label{whs} \feq
Any solution $R$ of \rref{whs} is clearly $C^1$. By derivation in $t$ of this equation, and
evaluation of the same at $t=t_0$, we get
\begin{prop}
\label{thc}
\textbf{Proposition.} If $u$ and $\EE$ are as in \rref{ee},
Eq.\rref{monodin} is equivalent to the problem
\beq {\dot R}(t) = U \ep(t) + U \ell(R(t), t) - B R(t)~, \qquad R(t_0) = U \delta~,
\label{thecon} \feq
for an unknown function $R \in C^1([t_0, t_1|, [0, \rho))$ (the terms
\textsl{control problem}, or \textsl{control equation} will be employed as well, for
\rref{thecon} or the differential equation therein).
\fine
\end{prop}
\section{Proof of Prop.\ref{main}.}
\label{prova}
We present in detail the argument in the case of a \textsl{compact} interval $[t_0, t_1]$. In this
case, we use the space $C([t_0, t_1], \FFF)$, regarded as a Banach space with the usual
sup norm $\| \psi \| := \max_{t \in [t_0, t_1]} \| \psi(t) \|$. The case when $\fiap$, $R$, etc.
are defined on $[t_0, t_1)$ (with $t_1$ possibly infinite) is treated in a similar way, using
$C([t_0, t_1), \FFF)$ with the topology of uniform convergence on all
compact subintervals $[t_0, \tau] \subset [t_0, t_1)$ ({\footnote{This
complete, locally convex topology on $C([t_0, t_1), \FFF)$ is defined by the
seminorms $\left(\|~\|_{\tau}\right)_{\tau \in [t_0, t_1)}$ where
$\| \psi \|_{\tau} := \max_{t \in [t_0, \tau]} \| \psi(t) \|$. To adapt the proof to this case,
the objects $\varrho$, $\overline{\TTT}(\fiap, \varrho)$, $\Lambda$, etc.
appearing in the sequel must be replaced by families of objects $\varrho_{\tau}$,
$\overline{\TTT}(\fiap \restriction [t_0, \tau], \varrho_\tau)$,
$\Lambda_{\tau}$, etc., one
for each $\tau$; the definition of $\DDD$ is simply rephrased using
$[t_0, t_1)$. The Peano-Picard iteration converges in all the seminorms $\|~\|_{\tau}$.}}).
Sticking from now on to the case $[t_0, t_1]$, we introduce the objects
\beq \varrho := \max_{t \in [t_0, t_1]} R(t)~; \qquad
\overline{\TTT}(\fiap, \varrho) = \{(f,t) \in \FFF \times [t_0,t_1]~|~
\| f - \fiap(t) \| \leq \varrho \} ~; \feq
\beq \DDD := \{ \psi \in C([t_0, t_1], \FFF)~|~\| \psi(t) - \fiap(t) \| \leq R(t) ~~\mbox{for $t
\in [t_0, t_1]$}~\}~; \feq
then $\varrho < \rho$, and $\overline{\TTT}(\fiap, \varrho)
\subset \TTT(\fiap, \rho)$ is a \textsl{strict} subset of $Dom \PP$. $\DDD$
is a closed subset of $C([t_0, t_1], \FFF)$ (containing $\fiap$) and
$\psi \in \DDD \Rightarrow \gr \psi \subset \overline{\TTT}(\fiap, \varrho)$.
\begin{prop}
\label{defj}
\textbf{Definition.} We put
\beq \JJ : \DDD \vain C([t_0, t_1], \FFF)~,
\qquad \psi \mapsto \JJ(\psi)~, \label{dej} \feq
$$ \hspace{1cm} \JJ(\psi)(t) := \UU(t - t_0) f_0 + \int_{t_0}^t ~d s~
\UU(t - s) \PP(\psi(s),s) \qquad \forall t \in [t_0, t_1]~. \hspace{2cm} \diamond $$
\end{prop}
Of course, we have
\begin{prop}
\textbf{Lemma.} $\varphi \in \DDD$
solves the Volterra problem \rref{int} if and only if $\varphi = \JJ(\varphi)$~. \fine
\end{prop}
\begin{prop}
\label{islip}
\textbf{Lemma.} There is a constant $\Lambda \geq 0$
such that, for all $\psi, \psi' \in \DDD$,
\beq \| \JJ(\psi)(t) - \JJ(\psi')(t) \| \leq \Lambda \int_{t_0}^t d s~ \| \psi(s) - \psi'(s) \|
\qquad \mbox{for $t \in [t_0, t_1]$}~. \label{verific} \feq
Thus $\| \JJ(\psi) - \JJ(\psi') \| \leq \Lambda (t_1 - t_0) \| \psi  - \psi' \|$,
which implies the continuity of $\JJ$.
\end{prop}
\textbf{Proof.} $\PP$ is Lipschitz at fixed time on the strict subsets of its domain, so there is a
constant $L \geq 0$ such that $\| \PP(f, t) - \PP(f', t) \| \leq L \| f - f' \|$ for $(f, t)$, $(f', t)$
$\in \overline{\TTT}(\fiap, \varrho)$.
If $u$ is an estimator for the semigroup $\UU$ and $U := \max_{t \in [t_0, t_1]} u(t)$,
we see that Eq.\rref{verific} is fulfilled with
$\Lambda := U L$; the remaining statements are trivial. \fine
\begin{prop}
\label{cent}
\textbf{Lemma.} $\JJ(\DDD) \subset \DDD$.
\end{prop}
\textbf{Proof.} Let $\psi \in \DDD$. For all $t \in [t_0, t_1]$
the definitions of $\JJ$ and of the error $E(\fiap)$, with the properties of
$\EE, u, \ell$, imply
\beq \JJ(\psi)(t) - \fiap(t) =  - E(\fiap)(t) + \int_{t_0}^t d s~\UU(t - s)~
[~\PP(\psi(s), s) - \PP(\fiap(s), s)~]~, \label{senzanor} \feq
\beq \| \JJ(\psi)(t) - \fiap(t) \| \leq \EE(t) + \int_{t_0}^t d s~u(t - s) \,\ell(\| \psi(s) - \fiap(s) \|, s)~.
\label{eqacc} \feq
On the other hand, $\| \psi(s) - \fiap(s) \| \leq R(s)$
which implies $\ell(\| \psi(s) - \fiap(s) \|, s) \leq \ell(R(s), s)$; inserting this into
\rref{eqacc}, and using the control inequality \rref{monod} for $R$, we conclude
\beq \hspace{4cm} \| \JJ(\psi)(t) - \fiap(t) \| \leq R(t)~, \qquad \mbox{i.e.},~~ \JJ(\psi) \in \DDD.
\hspace{1.5cm} \diamond \feq
The invariance of $\DDD$ under $\JJ$ is a central result; with the previously shown properties of $\JJ$, it
allows to set up the Peano-Picard iteration and get ultimately a fixed point.
\begin{defi}
\label{defik}
\textbf{Definition.} $(\varphi_k)$ $(k \in \naturali)$ is the sequence of
functions in $\DDD$, defined recursively by
\beq  \hspace{3cm} \varphi_0 := \fiap~, \qquad \varphi_k := \JJ(\varphi_{k-1}) \quad (k \geq 1)~.
\hspace{3cm} \diamond \feq
\end{defi}
\begin{prop}
\label{senzanom}
\textbf{Lemma.} For all $k \in \naturali$
and $t \in [t_0, t_1]$, it is
\beq \| \varphi_{k+1}(t) - \varphi_{k}(t) \| \leq \Sigma~{\Lambda^k (t - t_0)^k \over k!}
\label{num1} \feq
where $\Lambda$ is the constant of Eq.\rref{verific} and $\Sigma := \max_{t \in [t_0, t_1]} \EE(t)$. So,
\beq \| \varphi_{k + 1} - \varphi_{k} \| \leq \Sigma~
{\Lambda^k (t_1 - t_0)^k \over k!}~. \label{eqc} \feq
\end{prop}
\textbf{Proof.} Eq.\rref{eqc} is an obvious consequence of \rref{num1}. We will
prove \rref{num1} by recursion, indicating with
a subscript ${}_k$ the thesis at a specified order. \parn
We have $\varphi_{1}(t) - \varphi_0(t) =$ $\JJ(\fiap)(t) - \fiap(t) = - E(\fiap)(t)$ by
the definition of $E(\fiap)$, whence
$\| \varphi_{1}(t) - \varphi_{0}(t) \| \leq \EE(t) \leq \Sigma$;
this gives \rref{num1}$_0$. For each $k \geq 0$, we have
\beq \| \varphi_{k+2}(t) - \varphi_{k+1}(t) \| =
\| \JJ(\varphi_{k+1})(t) - \JJ(\varphi_{k})(t) \|
\leq \Lambda \int_{t_0}^ t d s \| \varphi_{k+1}(s) - \varphi_{k}(s) \|~, \label{sopra} \feq
the last passage depending on Eq.\rref{verific}. Eq.s \rref{sopra}
and \rref{num1}$_k$ imply \rref{num1}$_{k+1}$. \fine
\begin{lemma}
\label{legen}
\textbf{Lemma.} For all $k, k'$ and $n \in \naturali$, it is
\beq \| \varphi_{k'} - \varphi_k \| \leq \Sigma~e^{\Lambda (t_1- t_0)}~
{\Lambda^h (t_1 - t_0)^h \over h!}~,
\qquad h := \min(k, k')~;
\label{tesi} \feq
so, $(\varphi_k)$ is a Cauchy sequence.
\end{lemma}
\textbf{Proof.} To prove Eq.\rref{tesi}, it suffices to consider the case $k' > k$ (so that $h=k$).
Writing $\varphi_{k'} - \varphi_k =$ $\sum_{j=k}^{k'- 1}
(\varphi_{j+1} - \varphi_{j}) $ and using Eq.\rref{eqc} we get
\beq \| \varphi_{k'} - \varphi_k \| \leq \Sigma ~ \sum_{j=k}^{k' -1}
{\Lambda^j (t_1 - t_0)^j \over j!}~. \feq
On the other hand, for each $\xi \geq 0$, it is
$\sum_{j=k}^{k'-1} \xi^j/j! \leq$ $\sum_{j=k}^{+\infty} \xi^j/j! \leq$ $e^{\xi}~\xi^k/ k!$;~
with $\xi = \Lambda (t_1 - t_0)$ we obtain Eq.\rref{tesi}, implying
$\| \varphi_{k'} - \varphi_k \| \vain 0$ for $(k, k') \vain \infty$
({\footnote{Incidentally we note that \rref{tesi} could be improved, but this is
unnecessary: this estimate is needed only to infer the Cauchy property of the sequence.}). \fine
\vskip 0.2cm \noindent
\textbf{Proof of Prop.\ref{main}.} $(\varphi_k)$ being a Cauchy sequence,
$ \lim_{k \mapsto + \infty} \varphi_k := \varphi $ exists in $C([t_0, t_1], \FFF)$;
$\varphi$ belongs to $\DDD$ because this set is closed.
By the continuity of $\JJ$, we have
\beq \JJ(\varphi) = \lim_{k \mapsto + \infty} \JJ(\varphi_k) =
\lim_{k \mapsto + \infty} \varphi_{k+1} = \varphi~; \feq
thus, $\varphi$ solves the Volterra problem \rref{int}. Finally,
the inequality $\| \varphi(t) - \fiap(t) \| \leq R(t)$
for all $t \in [t_0, t_1]$ is ensured by the definition of $\DDD$.
\fine
\section{An elementary application of Prop.\ref{main}.}
\label{elem}
The results we are presenting in the forthcoming Prop.s \ref{prop1}, \ref{prop2}
are essentially known (see, e.g., \cite{Lak} for the case $\UU(t) = \bf{1}$ and
$\mbox{dim} \FFF$ finite), but their derivation as a subcase of Prop.\ref{main} is instructive: the main idea is to use the zero function as
an approximate solution.
\begin{prop}
\label{prop1}
\textbf{Proposition.} Consider the Volterra problem \rref{int}, where:
\parn
i) $\UU$ is a strongly
continuous linear semigroup, with an estimator $u(t) := U e^{-B t}$ ($U \geq 1$, $B \geq 0$).
\parn
ii) $\PP$ is continuous and Lipschitz at fixed time on the strict subsets of its open domain. It is
$\BBB(0, \rho) \times [t_0, T) \subset Dom \PP$ for some $\rho \in (0,+\infty]$, $T \in (t_0, +\infty]$, and
\beq \PP(0, t) = 0  \qquad \mbox{for $t \in [t_0, T)$.} \feq
There is a continuous function $\ell : [0, \rho) \times [t_0, T) \vain [0,+\infty)$, non decreasing in the
first variable, such that
\beq \| \PP(f, t) \| \leq \ell(\| f \|, t) \qquad
\mbox{for $(f, t) \in \BBB(0, \rho) \times [t_0, T)$.} \feq
iii) The control problem
\beq {\dot R}(t) = U \ell(R(t), t) - B R(t)~, \qquad R(t_0) = U \| f_0 \|, \label{caulel} \feq
has a solution $R \in C^1([t_0, t_N), [0, \rho))$, for some $t_N \in (t_0, T)$.  \parn
Then, the Volterra problem \rref{int} has a solution $\varphi \in C([t_0, t_N), \FFF)$ and,
for all $t$ in this interval,
\beq \| \varphi(t) \| \leq R(t)~. \feq
\end{prop}
\textbf{Proof.} We apply Prop.s \ref{main}, \ref{thc} with $\fiap(t) := 0$
for $t \in [t_0, t_N)$; the function $\ell$ in item ii) is a growth estimator for $\PP$ from
the approximate solution. The datum and differential errors are $d(\fiap) = - f_0$,
$e(\fiap)(t) = 0$, so they admit the estimators $\delta := \| f_0 \|$, $\ep(t) := 0$.
With these estimators, problem \rref{thecon} takes the form \rref{caulel}. \fine
The symbol $t_N$ adopted here for the right extreme of the domain of $R$ is chosen for
future convenience; it emphasizes the dependence of this object on the \textsl{norm} of the initial datum.
In the time independent case $\ell(r, t) = \ell(r)$, Eq.\rref{caulel} can be solved
by the quadrature formula
\beq \int_{U \| f_0 \|}^{R(t)} {d r \over U \ell(r) - B r} = t - t_0~; \label{quaf} \feq
let us write the explicit solution in a simple case.
\parn
\begin{prop}
\label{prop2}
\textbf{Proposition.} Let the previous assumptions be satisfied with $t_0 = 0$, $\delta=+\infty$, $T =+\infty$ and
$\ell(r, t) = P r^p$ $(P \geq 0,~ p > 1)$. Then, the problem \rref{caulel} has the
solution $R \in C^1([0, \tN), [0,+\infty))$ defined hereafter. It is
\beq \tN : =
\left\{ \barray{ll} + \infty & \mbox{if~~ $P U^p \| f_0 \|^{p-1} \leq B,$}
\\ {\displaystyle {{1 \over (p-1)} L_B \left(P U^p \| f_0 \|^{p-1} \right) }} &
\mbox{if~~ $P U^p \| f_0 \|^{p-1} > B, $} \farray~\right.
\label{t1} \feq
\beq L_B(u) := \left\{ \barray{ll} {\displaystyle - (1 / B) \log(1 - B/u )} &
\mbox{~~if $0 < B < u$,}
\\ 1 / u & \mbox{~~if $B =0 < u$;} \farray~\right. \feq
\beq R(t) := {U \| f_0 \| \over \left[~ 1 - ( P U^p \| f_0 \|^{p-1} - B )
E_{B} ((p-1) t )~ \right]^{1 \over p-1}} \qquad \mbox{for all $t \in [0, \tN)$}, \label{ft} \feq
\beq E_B(u) := \left\{ \barray{ll} {\displaystyle {e^{B u} - 1 \over B} } & \mbox{~~if $B > 0$,}
\\ u  & \mbox{~~if $B =0$.} \farray~\right. \feq
The function $R$ has the following features. If $P U^p \| f_0 \|^{p-1} < B$, $R$ is decreasing and
$R(t) \vain 0$ for $t \vain +\infty$. If $P U^p \| f_0 \|^{p-1} = B$, $R(t)=$ const.
$= U \| f_0 \|$. If
$P U^p \| f_0 \|^{p-1} > B$, $R$ is increasing and $R(t) \vain +\infty$ for $t \vain t^{-}_N$.
\end{prop}
\textbf{Proof.} Everything follows in an elementary way from \rref{quaf}. \fine
\textbf{Remarks.} i) A map $\PP$ as in the Example of page \pageref{exa} has the
properties required by the previous Proposition, if the function $t
\mapsto P(t)$ appearing in Eq.\rref{assu} is bounded on the interval $[t_0, T)$ under
consideration. In this case, the growth of $\PP$ from zero
admits the estimator $\ell(r, t) := P r^p$, with $P$ the sup of the function $t \mapsto P(t)$. \parn
ii) Obviously enough: if $P U^p \| f_0 \|^{p-1} < B$, the Volterra problem
\rref{int} has a solution $\varphi$ defined for all $t \in [0,+\infty)$, and
$\| \varphi(t) \| \leq R(t) \vain 0$ for $t \vain +\infty$. If $P U^p \| f_0 \|^{p-1} = B$,
we have again a solution defined on $[0, +\infty)$, and $\| \varphi(t) \| \leq U \| f_0 \|$ for all $t$.
If $P U^p \| f_0 \|^{p-1} > B$, we can grant existence of a solution $\varphi$ at least
until the time $\tN$ in Eq.\rref{t1}, and the bound $\| \varphi(t) \| \leq R(t)$ with $R$
diverging at $\tN$; the result for this case can be applied to blow up problems, to get a lower bound on
the time of explosion of the solution and an upper bound on its growth.
\parn
\textbf{Example.} We consider the Banach space
\beq \FFF := C_0(\reali) := \{ f : \reali \vain \reali~|~\mbox{$f$ is continuous, $f(x)
\vain 0$ for $x \vain \infty$} \}~; \feq
\beq \| f \| := \sup_{x \in \reali} | f(x) | \qquad \mbox{for $f \in \FFF$}. \feq
We define a linear semigroup $\UU : t \in [0,+\infty) \vain \UU(t) \in \LL(\FFF)$ setting
\beq \left(\UU(t) f \right)(x) := f(x+t) \feq
for all $f \in \FFF$; $\UU$ is strongly continuous, and $\| \UU(t) f \| = \| f \|$
({\footnote{In fact, $\UU$ can be extended to a linear
group, also defined for $t \leq 0$, but we do not emphasize this aspect: our general
framework is designed for time evolution in the future.}}). The generator
of $\UU$ is the operator
\beq \AA := {d \over d x} : C^1_0(\reali) \subset \FFF \vain \FFF ~, \qquad f \mapsto f_x \feq
where $C^1_0(\reali)$ is the space of the $C^1$ functions $f : \reali \vain \reali$ such that
$f, f_x$ vanish at infinity. \parn
We also introduce the function
\beq \PP : \FFF \vain \FFF~, \qquad \PP(f) := f^p \qquad \mbox{($p>1$ integer)}, \feq
which can be seen as a $t$-independent case of the Example on page \pageref{exa}, with
$\P(f_1,...,f_p) := f_1 ... f_p$; of course $\| \P(f_1,...,f_p) \| \leq \| f_1 \| ... \| f_p \|$.
\parn
We consider the Volterra problem \rref{int} with
$t_0 := 0$, and an arbitrary initial datum $f_0 \in \FFF$; in the
special case $f_0 \in C^1_0(\reali)$, the corresponding Cauchy problem  is
\beq \dot{\varphi}(t) = \varphi(t)_{x} + \varphi(t)^p~, \qquad \varphi(0) = f_0~, \label{wave} \feq
involving a first order wave equation with polynomial nonlinearity.
The results of the last Prop.\ref{prop2} can be applied in this framework
with $U=1$, $B=0$ and $P=1$. For any $f_0 \in \FFF$,
this Proposition ensures existence of the solution $\varphi$ from time $0$ to
\beq \tN := {\displaystyle {1 \over (p-1) \| f_0 \|^{p-1}}}~, \label{to} \feq
(intending $\tN := +\infty$ if $f_0 = 0$), and gives the bound
\beq \| \varphi(t) \| \leq R(t)~, \qquad R(t) := {\| f_0 \| \over \left[~ 1 - (p-1) \| f_0 \|^{p-1} t
~ \right]^{1 \over p-1}} \label{feas} \feq
for all $t \in [0, t_N)$.
In this case, the accuracy of the estimates in Prop.\ref{prop2} can be checked in a very direct way, because
the maximal solution of \rref{int} is known; this is given by
\beq \varphi(t)(x) = {f_0(x+t)  \over \left[~ 1 - (p-1) f_0(x+t)^{p-1} t
~ \right]^{1 \over p-1}} \qquad \mbox{for $t \in [0, \vartheta)$,} \label{compu} \feq
\beq \vartheta := \sup \Big\{t >0~\Big|~(p-1) \sup_{x \in \reali} f_0(x)^{p-1}~t < 1 \Big\}. \feq
If $f_0 \in C^1_0(\reali)$, the above $\varphi$ also fulfils the Cauchy problem \rref{wave}: this
implies the full equivalence of the Volterra and Cauchy problems for such an $f_0$,
in spite of the fact that item ii) of Prop.\ref{integral} does not apply to the non reflexive
Banach space $C_0(\reali)$.
The following facts occur: \parn
i) for $p$ odd, or $p$ even and $\sup_{x} f_0(x) = \sup_{x} | f_0(x) |$,
$\vartheta$ equals the time $\tN$ in Eq.\rref{to};
thus, Prop.\ref{prop2} gives the best possible lower bound on the
existence time of the maximal solution. \parn
ii) For $p$ even and $0 < \sup_{x} f_0(x) < \sup_{x} | f_0(x) |$, it is $+ \infty > \vartheta > t_N $. \parn
iii) For $p$ even and $\sup_{x} f_0(x) \leq 0 < \sup_{x} | f_0(x) |$, it is
$+ \infty = \vartheta > t_N $.
\vskip 0.1cm \noindent
The accuracy of the growth estimate \rref{feas} is easily analysed by comparison with Eq.\rref{compu}.
\parn
We think that better results would arise in cases ii) iii) by suitably generalizing
the theory of approximate solutions to the framework of \textsl{ordered} Banach spaces \cite{Zei1};
this will be done elsewhere.
\section{Approximate solutions on finite-dimensional submanifolds of $\FFF$.}
\label{galerk}
Let us discuss  a general scheme to construct accurate approximate solutions, and
apply to it Prop.\ref{main} to get information on the exact solution;
a typical realization of this scheme is the Galerkin method, discussed in the sequel. \parn
\textbf{The framework.} From now on: $\UU$ is a strongly continuous linear semigroup
with generator $\AA$ and an estimator $u(t) := U e^{- B t}$
($U \geq 1$, $B \in \reali$); $\PP : Dom \PP \subset \FFF \times \reali
\vain \FFF$ is continuous and Lipschitz on the strict subsets of its open domain;
$(f_0, t_0) \in Dom \PP$. \parn
Our idea is
to construct an approximate solution $\fiap$ for the Volterra problem
\rref{int}, lying on a finite-dimensional (linear or nonlinear) submanifold of $\FFF$;
we assume the latter to be coordinatized by some real parameters $a^k$, labelled
by a finite set of indices $I$.
More precisely, we consider an injective $C^1$ map
\beq \GG : Dom \GG \subset \reali^I \vain \FFF~, \qquad a = (a^k)_{k \in I} \mapsto \GG(a)~, \feq
with open domain, such that the partial derivatives
\beq \partial_{k} \GG(a) \equiv {\partial \GG \over \partial a^k}(a) \in \FFF \qquad (k \in I) \feq
are linearly independent for all $a \in Dom \GG$; we regard $Im \GG$ as a multidimensional surface in
$\FFF$. We also suppose that
\beq Im \GG \subset Dom \AA~,~~Im \GG \times [t_0,T) \subset Dom \PP \feq
for some $T \in (t_0,+\infty]$, and ask $\GG$ to be continuous as a map to $Dom \AA$ with the graph norm.
The approximate solution we consider has the form
\beq \fiap(t) := \GG(a(t))~, \qquad a(~) \in C^1([t_0, t_1), Dom \GG)~, ~~t \mapsto a(t) \feq
where $[t_0, t_1) \subset [t_0, T)$, and $a(~)$ is a function determined in the sequel.
Clearly, the
datum and differential errors of $\fiap$ are
\beq d(\fiap) := \GG(a(t_0)) - f_0~, \feq
\beq e(\fiap)(t) := \partial_{k} \GG(a(t)) {\dot a}^{k}(t) - \AA \GG(a(t)) -
\PP(\GG(a(t)), t) \feq
(here and in the sequel, we employ the familiar Einstein's  summation convention on repeated indices).
We prescribe $a(~)$ to fulfil a Cauchy problem
$$ {\dot a}(t) = X(a(t), t)~, \qquad a(t_0) = a_0 $$
where $a_0$ is an initial datum, and $X : Dom \GG \subset \reali^I \vain \reali^I$
a continuous vector field; the criteria to fix $a_0$ and $X$ are discussed later.
\parn
For convenience, for all $a \in Dom \GG$, $\dot{a} \in \reali^I$, $t \in [t_0, T)$ we put
\beq \hat{\delta}(a) := \| \GG(a) - f_0 \|~, \qquad
\hat{\ep}(a, \dot{a},t) :=
\| \AA \GG(a) + \PP(\GG(a), t) - \partial_{k} \GG(a) {\dot a}^{k} \|
\label{hatde} \feq
\beq \hat{\ep}(a,t) := \hat{\ep}(a, X(a,t),t)~; \feq
then, the approximate solution $\fiap$ admits the datum and differential error estimators
\beq \delta := \hat{\delta}(a_0)~, \qquad \ep(t) := \hat{\ep}(a(t),t)~. \feq
To conclude, we assume there are $\rho \in (0,+\infty]$ and a continuous function
\beq \hat{\ell} : [0, \rho) \times Dom \GG \times [t_0, T) \vain [0,+\infty)~,
\qquad (r,a,t) \mapsto \hat{\ell}(r,a,t)~, \feq
non decreasing in the variable $r$, such that $a \in Dom \GG$, $t \in [t_0, T)$ and $\| f - \GG(a) \| < \rho$
imply $(f, t) \in Dom \PP$ and
\beq \| \PP(f,t) - \PP(\GG(a),t) \| \leq \hat{\ell}(\| f - \GG(a) \|,a,t)~. \feq
Then, the function
\beq \ell(r,t) := \hat{\ell}(r,a(t),t) \feq
is a growth estimator for $\PP$ from $\fiap$. The application of Prop.s
\ref{main}, \ref{thc} gives
\begin{prop}
\label{analit}
\textbf{Proposition.} Consider the equations
\beq \dot{a}(t) = X(a(t),t)~, \qquad a(t_0) = a_0~, \label{anal1} \feq
\beq \dot{R}(t) = U \hat{\ep}(a(t),t) + U \hat{\ell}(R(t), a(t), t) - B R(t)~,
\qquad R(t_0) = U \hat{\delta}(a_0)~, \label{anal2} \feq
for the unknowns $a(~) \in C^1([t_0, t_1), Dom \GG)$, $R \in C^1([t_0, t_1),
[0, \rho))$. If $(a(~), R)$ is a solution on some interval $[t_0, t_1)$ and
$\fiap(t) := \GG(a(t))$, then the Volterra problem \rref{int} has a solution
$\varphi$ on $[t_0, t_1)$, and $\| \varphi(t) - \fiap(t) \| \leq R(t)$ on the same
interval.
\fine
\end{prop}
Let us pass to the criteria for choosing $X$ and $a_0$. One of the most familiar
is the Galerkin criterion (see, e.g., \cite{Dur} or \cite{Tem}): we will concentrate on it and will not discuss
other approaches
(such as the variational methods  often used for the Lagrangian or Hamiltonian evolution equations,
see, e.g., \cite{Kiv}).
The Galerkin choice for $a_0$ and $X$ is the one minimizing the norms of
the datum error and of the differential error (at any time):
\begin{prop}
\textbf{Definition.} The vector field $X$ and the datum $a_0$ fulfil the
Galerkin criterion if
\beq \hat{\ep}(a, \dot{a}, t) = \min! \qquad \mbox{for ${\dot a} = X(a, t)$,}
\label{gal2} \feq
\beq \hat{\delta}(a) = \min! \qquad \mbox{for $a=a_0$} \label{gal1} \feq
(the symbol $\min!$ indicating the absolute minimum).
\fine
\end{prop}
(Of course, condition \rref{gal1} is trivially satisfied if $f_0 = \GG(a_0)$;
then the absolute minimum of $\hat{\delta}$, attained at this point, is zero).
\parn
Both equations \rref{gal2} \rref{gal1} can be studied in a systematic way if
$\FFF$ is a Hilbert space, say real, with an inner product $<~|~>$ yielding the norm
$\| f \| := \sqrt{<f|f>}$; from now on we stick to this assumption. \parn
For all $a \in Dom \GG$ we
introduce the matrix
\beq \mat_{k l}(a) := < \partial_{k} \GG(a)\,|\,\partial_{l} \GG(a) >
\qquad (k, l \in I)~, \feq
which is symmetric and positive defined (recall the linear independence of the
vectors $\partial_{k} \GG(a)$). As customary in tensor calculus, we denote
the inverse matrix with $\mat^{k h}(a)$ and introduce the convention of "raising and lowering indices"
with these matrices. In connection with this, it is worthy to write down the identities
\beq < \partial^k \GG(a) \,| \,\partial_{h} \GG(a) > = \delta^{k}_{~h}~; \feq
$$ v^k(a) \partial_{k} \GG(a) = v_k(a) \partial^k \GG(a)~, $$
\beq < v^k(a) \partial_{k} \GG(a) \,|\,w^h(a) \partial_h \GG(a) > = v_k(a) w^k(a) = v^k(a) w_k(a)~. \feq
Here: $\partial^k G(a) := \mat^{k h}(a) \partial_h G(a) \in \FFF$; $v^k(a)$ ($k \in I$) is a family
of real numbers, $v_k(a) := \mat_{k h}(a) v^{h}(a)$ and $w^k(a)$, $w_k(a)$ have a similar
meaning.
We apply all these notations to the discussion of the minimum problems \rref{gal2}, \rref{gal1}; the
solutions are given by the two forthcoming Propositions.
\begin{prop}
\label{minimi2}
\textbf{Proposition.} For any fixed $(a, t) \in Dom \GG \times [t_0, T)$,
the function $\dot{a} \mapsto \hat{\ep}^2(a, \dot{a}, t)$
is quadratic; it has a unique point of absolute minimum at
\beq \dot{a}^{k} = X^{k}(a, t)~, \qquad
X^k(a,t) := < \partial^{k} \GG(a) \,|\, \AA \GG(a) + \PP(\GG(a), t) >~. \label{eqx} \feq
The absolute minimum $\hat{\ep}(a, X(a,t),t) := \hat{\ep}(a,t)$ is given by
$$ \hat{\ep}(a,t)^2 = \| \AA \GG(a) \|^2
- < \AA \GG(a) \,| \, \partial_k \GG(a) > < \partial^k \GG(a) \,| \,\ \AA \GG(a)> + $$
\beq + 2 < \AA \GG(a) \,| \,\PP(\GG(a), t) >
- 2 < \AA \GG(a) \,|\, \partial_k \GG(a) > < \partial^k \GG(a) \,| \,\PP(\GG(a), t) > + \label{eqep} \feq
$$ + \| \PP(\GG(a), t) \|^2
- < \PP(\GG(a),t) \,|  \,\partial_k \GG(a) > < \partial^k \GG(a) \,| \,\PP(\GG(a), t) >~. $$
\end{prop}
\textbf{Proof.} The general theory of Hilbert spaces
tells us that, given a vector $g$ and a closed vector subspace $\TT$ of $\FFF$, the problem
\beq \mbox{find $f$ in $\TT$ such that $\| g - f \| = \min!$} \label{hilb1} \feq
has the unique solution~
\beq f = \Pi g;~\| g - f \|^2 = \| g \|^2 - < g \,|\, \Pi g >,~
\mbox{$\Pi :=$ the orthogonal projection of $\FFF \vain \TT$}. \label{hilb2} \feq
For any fixed $(a, t)$, the problem we have in mind has just the form
\rref{hilb1}: in this case
\beq \TT = \TT(a) := \{ {\dot a}^k \partial_{k} \GG(a))~|~{\dot a} \in \reali^I \} \subset \FFF~,
\label{ta} \feq
(which represents the tangent subspace at $\GG(a)$ of $Im \GG$),
the unknown $f$ is written $X^{k}(a, t) \partial_{k} \GG(a)$ and
\beq g = \AA \GG(a) + \PP(\GG(a), t)~, \qquad
\qquad \Pi = \Pi(a) = <\partial^{k} \GG(a) \,|\, \boma{\cdot} >  \partial_{k} \GG(a)~; \label{themap} \feq
computing from \rref{hilb2} the solution $f$ and the quantity
$\| g - f \|^2 = \hat{\ep}(a,t)^2$, we obtain Eq.s \rref{eqx} \rref{eqep}.
\fine
\begin{prop}
\label{minimi1}
\textbf{Proposition}. Assume that: \parn
i) $\GG$ is $C^2$, $Dom \GG$ is convex; \parn
ii) the matrix $\mat_{k l}(a) +$ $< \GG(a) - f_0\,|\,\partial^2_{k l} \GG(a) >$
is semipositive for all $a \in Dom \GG$ (where $\partial^2_{k l}$ are the second
partial derivatives w.r.t. $a^{k}, a^{l}$); \parn
iii) there is a point $a_0$ such that
\beq < \partial^{k} \GG(a_0)\,|\,\GG(a_0) > = < \partial^{k} \GG(a_0)\,|\,f_0>~. \label{eqa0} \feq
Then, the absolute minimum of $\hat{\delta}(a)$ is attained at $a=a_0$.
\end{prop}
\textbf{Proof.} Everything follows computing the first and second derivatives of the
function $a \vain \hat{\delta}^{\,2}(a)$ in Eq.\rref{hatde}. Eq.\rref{eqa0} is the vanishing condition
for the first derivatives, the other assumptions ensure that the stationary point
$a_0$ is of absolute minimum. \fine
\textbf{Remark.} The geometrical meaning of Eq.\rref{eqa0} is $\Pi(a_0) \GG(a_0) = \Pi(a_0) f_0$,
where $\Pi(a_0)$ is the orthogonal projection onto $\TT(a_0)$, see \rref{themap}. \fine
\vskip 0.2cm \noindent
\textbf{The classical Galerkin method.}
All the previous formulas become very simple in the "classical" realization, where
\beq \GG : \reali^I \vain \FFF,~~ a \mapsto \GG(a) = a^{k} e_{k}~, \label{classical} \feq
$$ \mbox{$(e_{k})_{k \in I}$ linearly independent vectors of $Dom \AA$},~~
(e_k, t) \in Dom \PP~\mbox{for $k \in I, t \in [t_0, T)$}~. $$
In this case $Im \GG$ is a vector subspace, and we have the identities
\beq \partial_{k} \GG(a) = \mbox{const.} = e_{k},~~ \mat_{k l}(a) =
\mbox{const.} := \mat_{k l},~~\partial^{k} \GG(a) = \mbox{const.} = \mat^{k h} e_{h} := e^{k}~. \feq
(Also, the tangent space at any point $\GG(a)$ is $\TT(a) = \mbox{const.} = Im \GG$;
further simplifications occur in the orthonormal case where
$\mat_{k l} = \delta_{k l}$, $e^{k}= e_{k}$). \parn
\begin{prop}
\label{cl1}
\textbf{Proposition.} The absolute minimum point for $\hat{\delta}$ and its value at this point are
given by
\beq a^{k}_0 = < e^{k} \,|\, f_0 >~, \qquad \hat{\delta}(a_0) = \| f_0 - a^{k}_0 e_{k} \|~.
\label{eq3} \feq
(In particular, $\hat{\delta}(a_0) = 0$ if $f_0$ is in the linear subspace
spanned by the family $(e_k)$).
\end{prop}
\textbf{Proof.} Elementary (recall Eq.\rref{eqa0}). \fine
\begin{prop}
\label{cl2}
\textbf{Proposition.} Suppose $\PP$ to be
determined by a $p$-linear map $\P$, as in the Example of page \pageref{exa}. Then,
Eq.s \rref{eqx} and \rref{eqep} become
\beq X^{k}(a, t) = <e^k \,|\, \AA e_{l}> a^{l} + <e^k \,|\, \P(e_{l_1}, ..., e_{l_p}, t) > a^{l_1} ... a^{l_p} ; \label{eq1}
\feq
$$ \hat{\ep}(a,t)^2 = \Big(< \AA e_j \,|\, \AA e_l >  - < \AA e_j \,|\, e_k > < e^k \,|\, \AA e_l> \Big) a^j a^l + $$
\beq + 2 \Big( < \AA e_j \,|\, \P(e_{l_1},...,e_{l_p},t) >
- < \AA e_j | e_k > < e^k | \P(e_{l_1},...,e_{l_p},t) > \Big) a^j a^{l_1} ... a^{l_p} + \label{eq2} \feq
$$ + \Big( < \P(e_{j_1} ,..., e_{j_p},t) \,|\, \P(e_{l_1},...,e_{l_p},t) > + $$
$$ - < \P(e_{j_1} ,..., e_{j_p},t) \,|\, e_k> < e^k \,|\, \P(e_{l_1},...,e_{l_p},t) > \Big)
a^{j_1} ... a^{j_p} a^{l_1} ... a^{l_p}~. $$
In the r.h.s. of the last equation, the coefficients of $a^j a^l$ and $a^j a^{l_1} ... a^{l_p}$
are zero if the subspace spanned by the family $(e_k)$ is invariant under $\AA$ (which occurs, in particular,
if each $e_k$ is an eigenvector of $\AA$).
\end{prop}
\textbf{Proof. } Both Eq.s (\ref{eq1}-\ref{eq2}) follow easily from $\partial_k \GG(a) = e_k$
and from the multilinearity of $\P$. In
the second equation the coefficients of $a^j a^l$ and $a^j a^{l_1} ... a^{l_p}$ are, respectively,
\beq < \AA e_j \,|\, \AA e_l >  - < \AA e_j \,|\, e_k > < e^k \,|\, \AA e_l> = < \AA e_j \,|\, \AA e_l> -
<\AA e_j \,|\, \Pi \AA e_l>~, \feq
$$ < \AA e_j \,|\, \P(e_{l_1},...,e_{l_p},t) >
- < \AA e_j \,|\, e_k > < e^k \,|\, \P(e_{l_1},...,e_{l_p},t) > = $$
$$ = < \AA e_j \,|\, \P(e_{l_1},...,e_{l_p},t) > - < \Pi \AA e_j \,|\, \P(e_{l_1},...,e_{l_p},t) >~, $$
where $\Pi$ is the projection on the linear subspace spanned by $(e_k)$. If this
subspace is left invariant by $\AA$ we have $\Pi \AA e_l= \AA e_l$ for each $l \in I$, so the above coefficients are
zero. \fine
\begin{prop}
\label{elcl}
\textbf{Proposition.} Assume again $\PP$ to be as in the Example of page \pageref{exa}.
Then, the  growth of the function $\PP$ starting from a point
$\GG(a) = a^k e_k$ admits this estimate, for all $f \in \FFF$ and $t \in \Delta$:
\beq \| \PP(f,t) - \PP(\GG(a),t) \| \leq \hat{\ell}(\| f - \GG(a) \|, a,t) \label{eq4} \feq
\beq \hat{\ell} : [0,+\infty) \times \reali^I \times \Delta \vain [0,+\infty),~~
\hat{\ell}(r, a, t) := P(t) \sum_{j=1}^p \left( \barray{c} p \\ j \farray \right)
\sqrt{a_k a^k}^{~p-j} r^j~. \label{eq5} \feq
\end{prop}
\textbf{Proof.} Apply Eq.\rref{zero} with $f' = \GG(a)$ and $t' = t$; note that $\| f' \| =
\sqrt{a_k a^k}$. \fine
In applications of the classical Galerkin method, some choices for the family $(e_k)$
have a special consideration. Apart from systems of eigenvectors of $\AA$,
other choices occur in finite elements methods, which are
strictly related to the idea of approximating the evolutionary problem by space discretization; in this
case, $(e_k)$ is typically a family of piecewise linear (or polynomial)
"chapeau functions" related to some spatial grid, see e.g. \cite{Dur}.
\vskip 0.2cm \noindent
\section{Applications to the nonlinear heat equation.}
\label{appl}
Our aim is to discuss the nonlinear heat equation $\dot{f} = f_{x x} + f^p$ with Dirichlet boundary
conditions, for $x$ ranging in $(0,\pi)$ (we work in one space dimension only for
simplicity). Let us introduce
this equation in the framework of Sobolev spaces.
\vskip0.2cm\noindent
\textbf{Notations for Sobolev spaces.} All functions on $(0,\pi)$ are real-valued; $\FF(0,\pi)$ means
$\FF((0,\pi),\reali)$ for each functional class $\FF$.
We consider the Hilbert space $L^2(0,\pi)$, with
the standard inner product $<f\,|\,g>_{\! L^2} := \int_{0}^{\pi}~ d x f(x) g(x)$; here the functions
\beq s_k(x) := \sqrt{{2 \over \pi}}~ \sin(k x)  \qquad (k \in \{1,2,3,...\}) \label{sk} \feq
form a complete orthonormal system. We introduce the Sobolev space
\beq H^{1}(0,\pi) := \{ f \in L^2(0,\pi)~|~f_x \in L^2(0,\pi) \} \subset C([0,\pi])~, \label{incl} \feq
$f_x$ denoting the distributional derivative of $f$; this is a Hilbert space with the
inner product
\beq <f\,|\,g> := <f\,|\,g>_{\! L^2} + <f_x\,|\,g_x>_{\! L^2}~, \label{inner} \feq
yielding the norm $\| f \| := \sqrt{<f\,|\,f>}$. The inclusion indicated in
\rref{incl} is well known, and allows to define
$f(0), f(\pi)$ for all $f \in H^1(0,\pi)$; we fix the attention on the closed subspace
\beq \FFF := H^{1}_0(0,\pi) := \{ f \in H^1(0,\pi)~|~f(0) = f(\pi) = 0 \}~, \feq
and equip it with the restriction of the inner product
\rref{inner}. It turns out that
\beq \FFF = \{ f \in L^2(0,\pi)~|~\sum_{k=1}^{\infty} k^2 <s_k\,|\,f>^{2}_{\! L^2}\, < +\infty \}~. \feq
The functions $s_k$ form a complete orthogonal system for this space: it is
$$ < s_k \,|\, s_l > = ({1 + k^2}) \delta_{k l}, \qquad <s_k | f > = (1 + k^2) < s_k | f >_{\! L^2}
~\forall f \in \FFF~, $$
\beq <f\,|\,g> = \sum_{k=1}^{\infty} (1 + k^2) <f\,|\,s_k>_{\! L^2} <s_k\,|\,g>_{\! L^2}~
\forall f,g \in \FFF. \feq
Both $H^1(0,\pi)$ and $\FFF$ are known to be Banach algebras with respect to the
pointwise product. We will use systematically the inequality
(almost optimal, see the Appendix \ref{appe1})
\beq \| f g \| \leq \| f \| \, \| g \| \qquad \forall f, g \in \FFF~. \label{multip1} \feq
\vskip 0.2cm \noindent
\textbf{The operator $\boma{\AA}$}. This is the linear map
\beq \AA := {d^2 \over d x^2}~~\mbox{on}~~Dom \AA := \{ f \in \FFF~|~f_{xx} \in \FFF \}~. \feq
Of course, for all $k$,
\beq \AA s_k = -k^2 s_k~. \feq
The operator $\AA$ generates the strongly continuous linear semigroup $\UU$ on $\FFF$, defined by
\beq \UU(t) f  := \sum_{k=1}^{\infty} e^{-k^2 t} <s_k\,|\,f>_{\! L^2} s_k
\qquad \mbox{for $t \in [0,+\infty)$, $f \in \FFF$}; \label{semiheat} \feq
the above series is in fact convergent in $\FFF$, and
\beq \| \UU(t) f \| = \sqrt{ \sum_{k =1}^{\infty} e^{- 2 k^2 t} (1 + k^2)
<s_k\,|\,f>^2_{\! L^2} } \,\leq \,e^{- t} \| f \|~. \feq
\vskip 0.2cm \noindent
\textbf{The nonlinear function $\boma{\PP}$}. This is defined by
\beq \PP : \FFF \vain \FFF~, \qquad \PP(f) := f^p \qquad \mbox{($p > 1$ integer).} \feq
It belongs to the class of maps in the Example of page \pageref{exa}, and corresponds
to the time independent $p$-linear map
\beq \P : \times^p \FFF \vain \FFF~, \qquad \P(f_1,...,f_p) := f_1 ... f_p~. \label{dp} \feq
Of course, Eq.\rref{multip1} implies $\| \P(f_1, ..., f_p) \| \leq \| f_1 \| ... \| f_p \|$.
\vskip 0.2cm \noindent
\textbf{The Volterra and Cauchy problems.} We consider the Volterra problem \rref{int} on $\FFF$,
with $\UU$ as before, $\PP(f,t) := $ the above defined $\PP(f)$, $t_0 :=0$ and some initial
datum $f_0 \in \FFF$; this reads
\beq \varphi(t) = \UU(t) f_0 + \int_{0}^t d s~ \UU(t-s) \varphi(s)^p~. \label{eqvol} \feq
From now on, we will denote with
\beq \varphi : [0, \vartheta) \vain \FFF \feq
the maximal solution. If $f_0 \in Dom \AA$, \rref{int} is fully equivalent to the Cauchy problem
\beq {\dot \varphi}(t) = \varphi(t)_{x x} + \varphi(t)^p~, \qquad \varphi(0) = f_0 \label{eqcau} \feq
(because $\FFF$ is reflexive and $\PP$ Lipschitz on the strict subsets of its domain).
\vskip0.2cm\noindent
\textbf{Kaplan's blow up criterion.} We specialize to the present
framework a general citerion of Kaplan \cite{Kap} for the blow up in a finite time of
the solution of a nonlinear parabolic equation. To this purpose, we introduce the function
\beq Q : L^2(0,\pi) \vain \reali \qquad  f \mapsto Q(f) := {1 \over 2}  <\sin | f >_{\! L^2} =
{1 \over 2} \int_{0}^{\pi} d x~\sin x \,f(x)~. \label{deq} \feq
\begin{prop}
\label{kapl}
\textbf{Proposition.} Consider the Volterra problem \rref{eqvol}; if
\beq f_0 \geq 0~, \qquad Q(f_0) > 1~, \feq
then
\beq \vartheta \leq \TK~, \qquad \TK := - {1 \over (p-1)} \log \left(1 - {1 \over Q(f_0)^{p-1}}\right)~.
\label{det1} \feq
\end{prop}
\textbf{Proof.} It is sketched in the Appendix \ref{appe2}, adapting Kaplan's general argument. \fine
From now on, $\TK$ will be referred to as the \textsl{Kaplan time} for the datum $f_0$.
\vskip 0.2cm \noindent
\textbf{Basic estimates on the solution.} As a first step in our analysis, let us apply
Prop.s \ref{prop1} and \ref{prop2} to the
Volterra problem. In the case we are considering, the semigroup has
an estimator $u(t) = U e^{-B t}$ with $U=1$, $B=1$; also,
it is $\| \PP(f) \| \leq \ell(\| f \|)$ with $\ell(r) := r^p$. Therefore, we have
\begin{prop}
\label{basic}
\textbf{Proposition.} For any initial datum $f_0 \in \FFF$, it is $\vartheta \geq \tN$ and
$\| \varphi(t) \| \leq R(t)$ for all $t \in [0,\tN)$, with $\tN$ and $R$ depending
on the norm of $f_0$ in the following way:
\beq \tN : = \left\{ \barray{ll} + \infty & \mbox{if~~ $\| f_0 \| \leq 1,$}
\\ {\displaystyle - {1 \over (p-1)} \log \left(1  - {1 \over {\| f_0 \|}^{p-1}} \right) } &
\mbox{if~~ $\| f_0 \|  > 1, $} \farray~\right.
\label{t1c} \feq
\beq R(t) := {\| f_0 \| \over \left[~ 1 - ( \| f_0 \|^{p-1} -1 )
(e^{(p-1) t} - 1)~ \right]^{1 \over p-1}}~. \label{ftc} \feq
If $\| f_0 \| < 1$, $R$ is decreasing and
$R(t) \vain 0$ for $t \vain +\infty$. If $\| f_0 \| = 1$, $R(t)=1$ for all $t$. If
$\| f_0 \| > 1$, $R$ is increasing and $R(t) \vain +\infty$ for $t \vain \tN^{-}$.
\fine
\end{prop}
\textbf{Summary of the previous results on $\boma{\vartheta}$. An example.} We have
\beq \tN \leq \vartheta~\mbox{for all $f_0 \in \FFF$}~; \qquad
\vartheta \leq \TK \quad \mbox{if $f_0 \geq 0$,~ $Q(f_0) > 1$}, \label{theb} \feq
with $\tN$ as in \rref{t1c}, $\TK$ as in \rref{det1}. Let us consider, in particular, the initial datum
\beq f_0(x) := A s_1(x) = \sqrt{2 \over \pi} A \sin x~, \qquad (A \geq 0)~; \label{f0} \feq
then $f_0 \in Dom \AA$, so we have a full equivalence of \rref{eqvol} with the Cauchy problem \rref{eqcau}.
It turns out that
\beq \| f_0 \| = {A \over \CN},~ \CN := {\sqrt{2} \over 2} = 0.7071..
 \quad ; \quad Q(f_0) = {A \over \CK},~ \CK := {2 \sqrt{2 \over \pi}} = 1.595..~. \feq
Therefore, Eq.\rref{theb} with this choice of the datum tells us that
\beq \vartheta  = +\infty~~\mbox{if~ $0 \leq A \leq \CN$};~~\tN  \leq \vartheta ~~
\mbox{if $\CN < A \leq \CK$};~~ \tN  \leq \vartheta  \leq \TK ~~ \mbox{if $A > \CK$}~, \label{estimates} \feq
\beq \tN  := - {1 \over p-1} \log \left(1 - {\CN^{p-1} \over A^{p-1}} \right)
\sim_{A \vain +\infty} {1 \over p-1}~{\CN^{p-1} \over A^{p-1}}~, \label{t1a} \feq
\beq \TK  := - {1 \over p-1} \log \left(1 - {\CK^{p-1} \over A^{p-1}} \right)
\sim_{A \vain +\infty} {1 \over p-1}~{\CK^{p-1} \over A^{p-1}}~. \label{t2a} \feq
It should be noted that \rref{theb} does not allow to establish whether $\vartheta $
is finite or infinite, for $A$ in the interval $(\CN, \CK]$.
In the rest of the Section, we will infer more precise estimates about $\vartheta $
by means of the Galerkin method, and also rediscuss its behaviour for large $A$.
\vskip 0.2cm \noindent
\textbf{A Galerkin approach to the nonlinear heat equation.} We apply the scheme
of Sect.\ref{galerk} with
\beq I~ \mbox{a finite subset of}~ \{1,2,3,....\}~, \qquad
\GG(a) := a^k s_k~ \mbox{for all $a = (a^k) \in \reali^I$} \feq
and $s_k$ the functions \rref{sk}. We refer, in particular, to the description
given in the previous Section for the "classical" Galerkin method,
to be employed with $e_k := s_k$,
$< | >$ the inner product \rref{inner} on $\FFF := H^1_{0}(0,\pi)$, and
\beq \mat_{k l} = (1 + k^2)~ \delta_{k l}~, \qquad \mat^{k l} = {1 \over 1 + k^2}~\delta^{k l}~; \feq
these matrices are used to raise and lower indices.
The vector field $X$, the error function $\hat{\epsilon}$ and
the growth estimator function $\hat{\ell}$ of Eq.s \rref{eq1} \rref{eq2} \rref{eq5}
are time independent,
and given by
\beq X^k(a) := - k^2 a^k + <s^k | s_{l_1} ... s_{l_p} > a^{l_1} ... a^{l_p}~, \label{xcase} \feq
\beq \hat{\epsilon}(a)^2 := (<s_{j_1} ... s_{j_p} | s_{l_1} ... s_{l_p} >
- <s_{j_1} ... s_{j_p} | s_k > <s^k | s_{l_1} ... s_{l_p} >) a^{j_1} ... a^{j_p} a^{l_1} ... a^{l_p}~,
\label{epcase} \feq
\beq \hat{\ell}(r, a) := \sum_{j=1}^p \left( \barray{c} p \\ j \farray \right)
\sqrt{a_k a^k}^{~p-j} r^j \label{whe} \feq
(see the observation following Eq.\rref{dp}; also, note that $a_k a^k = \sum_{k \in I} (1 + k^2) (a^k)^2$).
This amount of information (completed by Eq.\rref{eq3} for the initial datum)
must be inserted into the general scheme
of Prop.\ref{analit}; the solution of the finite
dimensional system \rref{anal1},\rref{anal2} appearing therein provides
simultaneously: \parn
i) a pair of functions $a(~)$, $R(~)$ on an interval $[0, \tG)$, the former
giving the approximate solution $\varphi_{ap}(t) := a^k(t) s_k$. In the sequel $\tG$
will be called the \textsl{Galerkin time}; \parn
ii) an assurance that the Volterra problem \rref{eqvol} has a maximal solution $\varphi$ on
$[0, \theta) \supset [0, \tG)$, and a bound
$\| \varphi(t) - \varphi_{ap}(t) \| \leq R(t)$ for all $t < \tG$.
\vskip 0.2cm \noindent
\textbf{Introducing an example.} We assume
\beq p := 2~, \qquad f_0~ \mbox{as in \rref{f0}}~. \feq
Problem (\ref{eqvol}-\ref{eqcau}) will be treated with a "two-modes"
application of the Galerkin method; more precisely, we will work on the linear
submanifold spanned by $(s_k)_{k\in I}$, setting
\beq I := \{1,3\}~, \qquad \alpha := a^1, \quad \gamma := a^3~; \feq
(the choice $I = \{1,2,3\}$ would not yield any improvement, because
the function $t \mapsto a^2(t)$ would be ultimately found to be zero).
The vector field, the error function and the growth function of Eq.s \rref{xcase}
\rref{epcase} \rref{whe} are given by
\beq X^\alpha(\alpha, \gamma)  = - \alpha + \sqrt{2 \over \pi^3}
\left({8 \over 3} \alpha^2 - {16 \over 15} \alpha \gamma
+ {72 \over 35} \gamma^2 \right)~, \feq
$$ X^\gamma(\alpha, \gamma) := - 9 \gamma + \sqrt{2 \over \pi^3}
\left(- {8 \over 15} \alpha^2 + {144 \over 35} \alpha \gamma
+ {8 \over 9} \gamma^2 \right)~; $$
\beq \hat{\ep}(\al, \gamma)^2 = \left( {7 \over 2 \pi} - {512 \over 15 \pi^3} \right) \alpha^4 +
\left( {34816 \over 315 \pi^3} - {10 \over \pi} \right) \alpha^3 \gamma + \feq
$$ + \left( {46 \over \pi} - {12172288 \over 33075 \pi^3} \right) \alpha^2 \gamma^2
- {22528 \over 175 \pi^3} \alpha \gamma^3 + \left( {39 \over 2 \pi} -
{3247616 \over 99225 \pi^3} \right) \gamma^4~; $$
\beq \hat{\ell}(\alpha, \gamma, r) = r^2 + 2 \sqrt{2 \alpha^2 + 10 \gamma^2}~ r~. \feq
According to \rref{eq3}, the initial conditions for $\alpha(t)$ and $\gamma(t)$ are, respectively,
\beq <s^1 \,|\, f_0> = A~, \qquad <s^3\, |\, f_0> = 0~; \feq
the corresponding datum error is zero. In conclusion, we have to study the system
\beq \dot{\alpha} = X^{\alpha}(\alpha, \gamma)~, \qquad
\dot{\gamma} = X^{\gamma}(\alpha, \gamma)~, \qquad
{\dot R} = \hat{\ep}(\alpha, \gamma) + \hat{\ell}(\alpha, \gamma, R) - R~, \label{conc} \feq
$$ \alpha(0) = A~, \qquad \gamma(0) = 0~, \qquad R(0) = 0~, $$
for the unknown functions $t \mapsto \alpha(t), \gamma(t), R(t)$. This cannot be solved
analytically, but can be easily treated by any package for the numerical solution
of ordinary differential equations; an integration algorithm with adaptative
control of the step size gives an excellent approximation
for the solution of \rref{conc}, also including the evaluation of its existence time.
All statements that follow about the system \rref{conc} are based
on the MATHEMATICA package; thus, expression such as "the solution of \rref{conc}", etc., always
indicate the MATHEMATICA output (of which we report the first digits).
\vskip0.2cm\noindent
\textbf{New estimates on the existence time $\boma{\vartheta}$.} We have the bounds
\beq \tG  \leq \vartheta ~\mbox{for all $A \geq 0$}, \qquad \vartheta  \leq \TK ~\mbox{for
$A > \CK$} \feq
involving the Galerkin and Kaplan times, defined previously (the latter depends on $\CK =1.595..$).
It is found that
\beq \tG  = +\infty~~\mbox{(whence $\vartheta  = +\infty$)} \qquad \mbox{for $0 \leq A \leq \CCG$}~,
~~\CCG = 1.056..~; \feq
\beq \alpha(t), \gamma(t), R(t) \vain 0 \qquad \mbox{for $t \vain +\infty$~ and~ $0 \leq A \leq \CCG$.} \feq
It should be noted that the bound $\CCG$ on $A$ for the global existence of \rref{eqcau} is better
than the previously derived bound $\CN = 0.7071..$. For larger values of $A$, the existence
time $\tG $ for \rref{conc} is finite. The forthcoming Table reports $\tG $ for
some values of $A$ above the Kaplan critical value $\CK$, as well as the corresponding
values of $\TK $. It also reports
\beq \eta  := {\TK  - \tG  \over \TK  + \tG }~, \label{eta} \feq
which is the relative uncertainty about the actual existence time $\vartheta $ of
\rref{eqcau} corresponding to the lower and upper bounds $\tG $, $\TK $.
\beq
\begin{tabular}{|l|l|l|l|} \hline
$A$ & $\tG $ & $\TK $ & $\eta $ \\ \hline
$\leq 1.056..$ & $+\infty$ & $~$ & $~$ \\
$1.60$ & $1.104..$ & $5.935..$ & $0.6861..$ \\
$2$ & $0.7730..$ & $1.598..$ & $0.3481..$ \\
$4$ & $0.3138..$ & $0.5090..$ & $0.2372..$ \\
$10$ & $0.1112..$ & $0.1738..$ & $0.2196..$ \\
$20$ & $0.05340..$ & $0.08315..$ & $0.2177..$ \\ \hline
\end{tabular}
\feq
The $A \vain +\infty$ limit for the previous estimates is easily discussed.
To determine the asymptotics of $\tG $, we
reexpress the unknown functions $\alpha(t)$, $\gamma(t)$ and $R(t)$
in terms of three rescaled functions $\ti \vain \ta(\ti), \tc(\ti), \tF(\ti)$, depending
on $\ti := A t$ and defined by
\beq \alpha(t) = A~ \ta(A t)~, \qquad \gamma(t) = A~ \tc(A t)~, \qquad R(t) = A~ \tF(A t)~. \feq
Then, the system \rref{conc} becomes (with $' := d/ d \ti$)
$$ \ta' = - {\ta \over A} + \sqrt{2 \over \pi^3}
\left({8 \over 3} \ta^2 - {16 \over 15} \ta \tc
+ {72 \over 35} \tc^2 \right),~ \tc' = - {9 \tc \over A} + \sqrt{2 \over \pi^3}
\left(- {8 \over 15} \ta^2 + {144 \over 35} \ta \tc
+ {8 \over 9} \tc^2 \right), $$
\beq \tF' = \hat{\ep}(\ta, \tc) + \hat{\ell}(\ta, \tc, \tF) - {\tF \over A} \label{conca} \feq
$$ \ta(0)= 1~, \qquad \tc(0) = 0~, \qquad \tF(0) = 0~. $$
In the $A \vain + \infty$ limit, the terms
$\ta/A$, $\tc/A$ and $\tF/A$ can be neglected and the outcoming system
$\mbox{\rref{conca}}_{\infty}$ is $A$-independent.
The numerical treatment of this limit system shows that the solution $\ti \vain \ta(\ti),
\tc(\ti), \tF(\ti)$ exists for $\ti \in [0,\CG)$, where
\beq \CG = 1.026..~. \feq
Returning to the standard time $t = \ti/A$, we conclude that
\beq \tG  \sim {\CG \over A}~~ \mbox{for $A \vain +\infty$}~. \label{atg} \feq
It should be noted that all values of $A$ in the previous Table are seen empirically
to fulfil the inequality
\beq \tG  \geq - {\CG \over \CCG} \log \left(1 - {\CCG \over A} \right)~, \feq
with $\tG $ very close to the r.h.s. To compare these results
with the Kaplan upper bound recall that, for all $A > \CK = 1.596..$,
\beq \TK  = - \log \left(1 - {\CK \over A} \right) \sim_{A \vain +\infty} {\CK \over A}~. \label{atk} \feq
Due to \rref{atg} \rref{atk}, the relative uncertainty \rref{eta} has the limit
\beq \eta  \vain {\CK-\CG \over \CK+\CG} = 0.2173.. \qquad \mbox{for $A \vain + \infty$.} \feq
As a matter of fact, in this limit case one can find directly
the asymptotics for the actual existence time $\vartheta$ of (\ref{eqvol}-\ref{eqcau}).
In fact, if one writes the maximal solution $\varphi$ as
\beq \varphi(t) = A \chi(A t) \feq
one obtains for $\chi$ the Cauchy problem
\beq \chi'(\ti) = {1 \over A} \chi(\ti)_{xx} + \chi^2(\ti)~, \qquad \chi(0)(x) = \sqrt{2 \over
\pi} \sin x~. \feq
For $A \vain +\infty$, the differential equation becomes $\chi'(\ti) = \chi^2(\ti)$, and the solution is
\beq \chi(\ti)(x) = {\sqrt{2/\pi} \sin(x) \over 1 - \sqrt{2/\pi} \, \ti \sin x} \qquad \mbox{for
$\ti \in [0, \sqrt{\pi/2})$}~; \feq
so, returning to the standard time $t = \ti/A$, we conclude
\beq \vartheta  \sim_{A \vain +\infty} {\sqrt{\pi/2} \over A} = {1.253.. \over A}~. \label{atheta} \feq
The constant $\sqrt{\pi/2}$ is fairly close to the
arithmetic mean of the costants $\CG$ and $\CK$, which appear in the asymptotic
expressions \rref{atg}, \rref{atk} for the lower and upper bounds $\tG $, $\TK $. Thus,
the actual existence time $\vartheta$ should be close to the arithmetic mean of $\tG $ and $\TK $ if
$A$ is sufficiently large. An attack to the Cauchy problem \rref{eqcau}
that we performed approximating $d^2/ d x^2$ by finite differences
seems to indicate that this actually occurs for all $A \gtrsim 2$.
\vskip0.2cm\noindent
\textbf{Analysis of the Galerkin solution.} After solving the system \rref{conc} for a given
$A$, one constructs the corresponding approximate solution for \rref{eqcau} such that,
for $t \in [0, \tG)$,
\beq \varphi_{ap}(t)(x) =
\alpha(t) \sqrt{2 \over \pi} \sin x + \gamma(t) \sqrt{2 \over \pi} \sin(3 x);
\quad \| \varphi_{ap}(t) \| = \sqrt{2 \alpha(t)^2 + 10 \gamma(t)^2}~. \feq
The system \rref{conc} also gives a function $R$ such that $\| \fiap(t) - \varphi(t) \| \leq R(t)$
for $t$ in the same interval;
let us illustrate the behaviour of the above functions for two values of $A$. \parn
\textbf{Case $\boma{A=1}$.} It is $\tG = +\infty$, which implies $\vartheta = +\infty$.
Figures \ref{f1}, \ref{f2}, \ref{f4} give the graphs of the functions $t \mapsto \alpha(t)$,
$\gamma(t)$,  $\| \fiap(t) \|$ and $R(t)$, all converging to zero for $t \vain +\infty$.
Figure \ref{f3}
gives the function $x \in (0,\pi) \mapsto \fiap(t)(x)$ at three fixed times.
For the exact solution $\varphi$ of \rref{eqcau}, we infer
\beq \| \varphi(t) \| \leq \| \fiap(t) \| + R(t) \vain 0 \qquad \mbox{for $t \vain +\infty$}~. \feq
Figure \ref{f5} is a
graph of the relative bound $R(t)/\| \fiap(t) \|$ in a time interval where it is fairly little. \parn
\textbf{Case $\boma{A=4}$.} The Galerkin system \rref{conc} has a finite
existence time $\tG = 0.3138..$. Figures \ref{f6}-\ref{f10} give information
of the same kind as the figures of the case $A=1$, but describe a qualitatively different behaviour;
in particular, the function $t \mapsto R(t)$ diverges for $t \vain \tG$.
\begin{figure}
\parbox{3in}{
\includegraphics[
height=2.0in,
width=2.8in
]%
{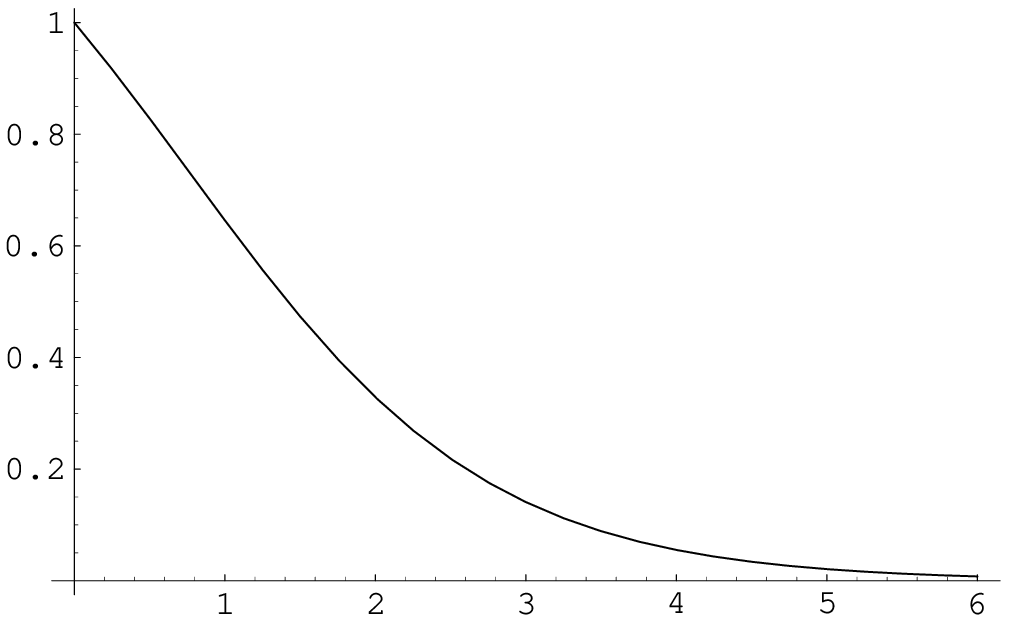}%
\caption{$A=1$. Graph of $\alpha(t)$.}
\label{f1}
}
\~\
\parbox{3in}{
\includegraphics[
height=2.0in,
width=2.8in
]%
{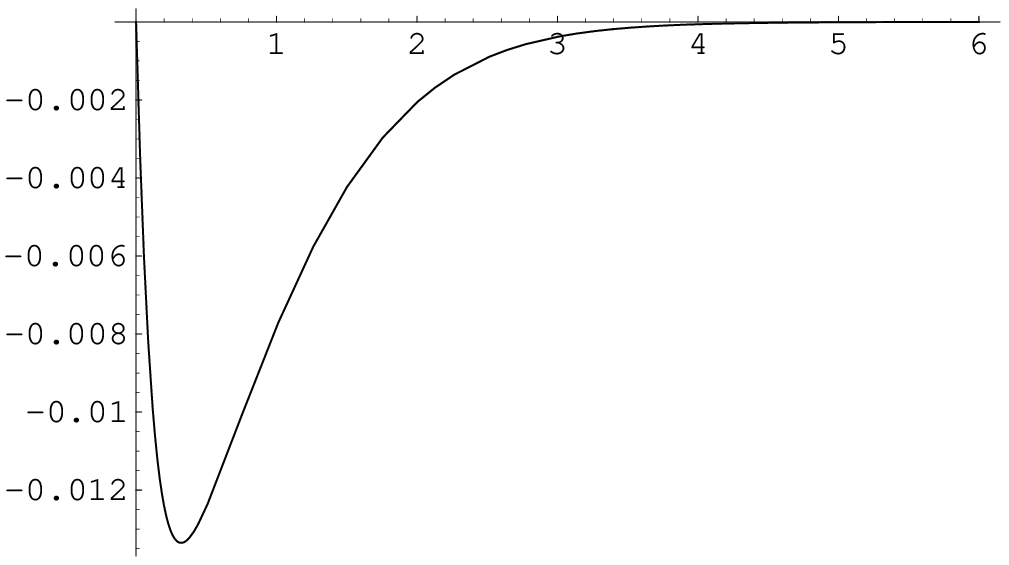}%
\caption{$A=1$. Graph of $\gamma(t)$.}
\label{f2}
}
\begin{center}
\includegraphics[
height=2.0in,
width=2.8in
]%
{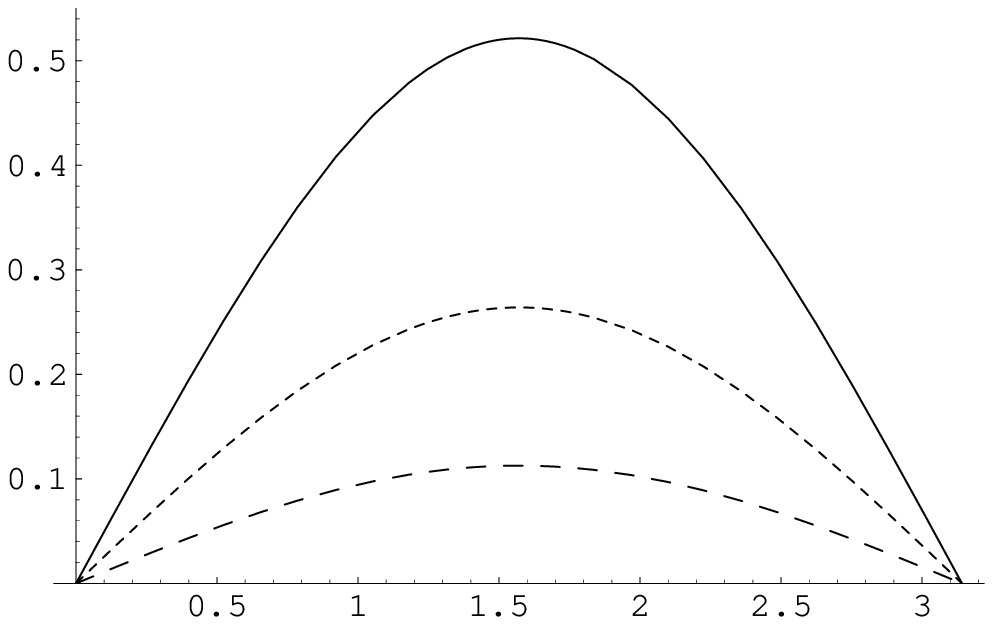}%
\caption{$A=1$. Graphs of $\fiap(t)(x)$ for $x \in (0, \pi)$ and $t=1$ (continuous line),
$t=2$ (short dashes), $t=3$ (long dashes).}
\label{f3}
\end{center}
\parbox{3in}{
\includegraphics[
height=2.0in,
width=2.8in
]%
{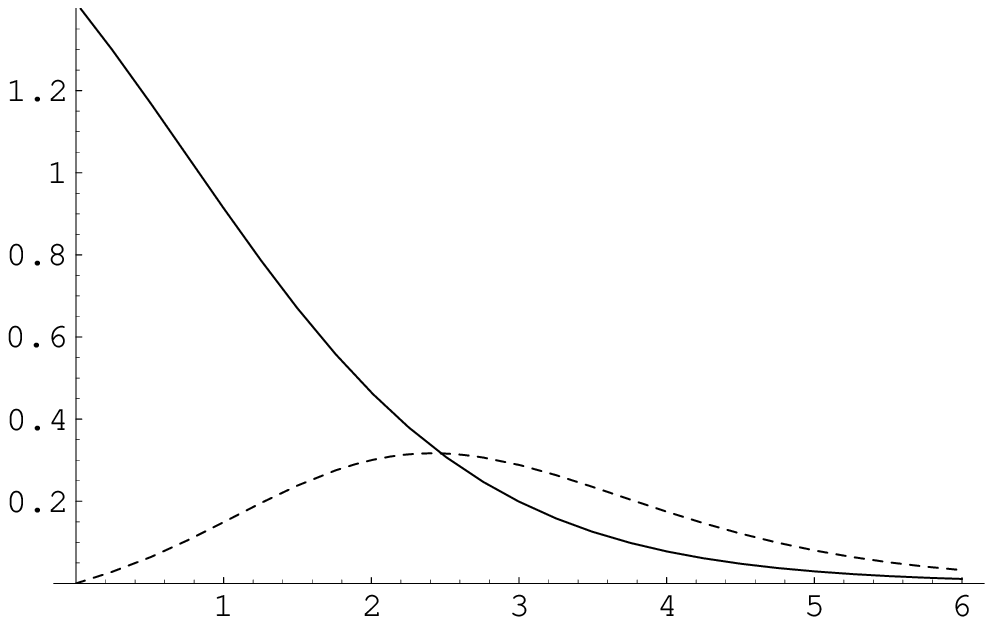}%
\caption{$A=1$. Graphs of $\| \fiap(t) \|$ (continuous line) and $R(t)$  (dashed line).}
\label{f4}
}
\~\
\parbox{3in}{
\includegraphics[
height=2.0in,
width=2.8in
]%
{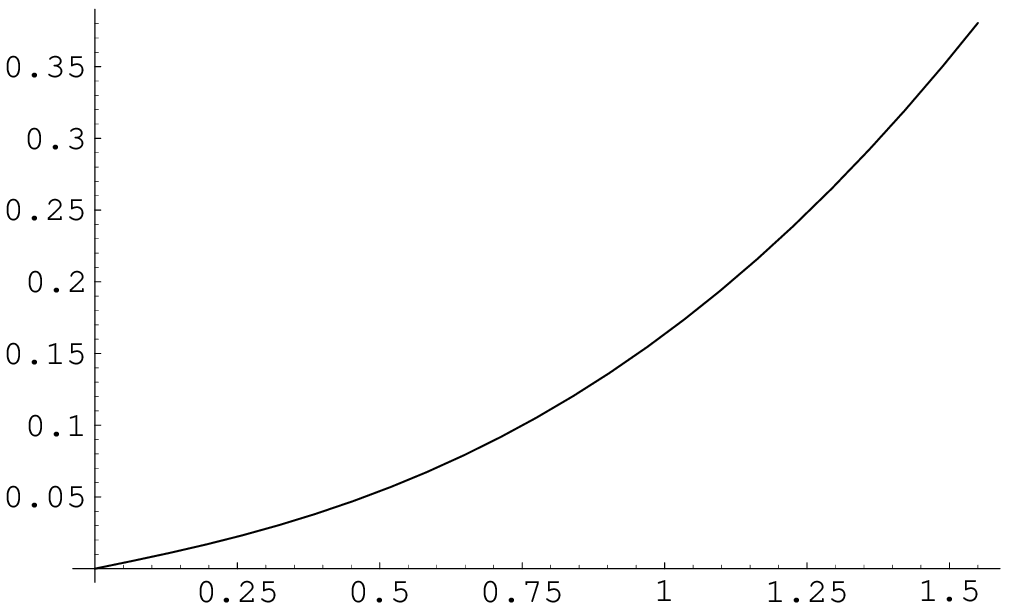}%
\caption{$A=1$. Graph of $R(t)/\| \fiap(t) \|$.}
\label{f5}
}
\end{figure}
\begin{figure}
\parbox{3in}{
\includegraphics[
height=2.0in,
width=2.8in
]%
{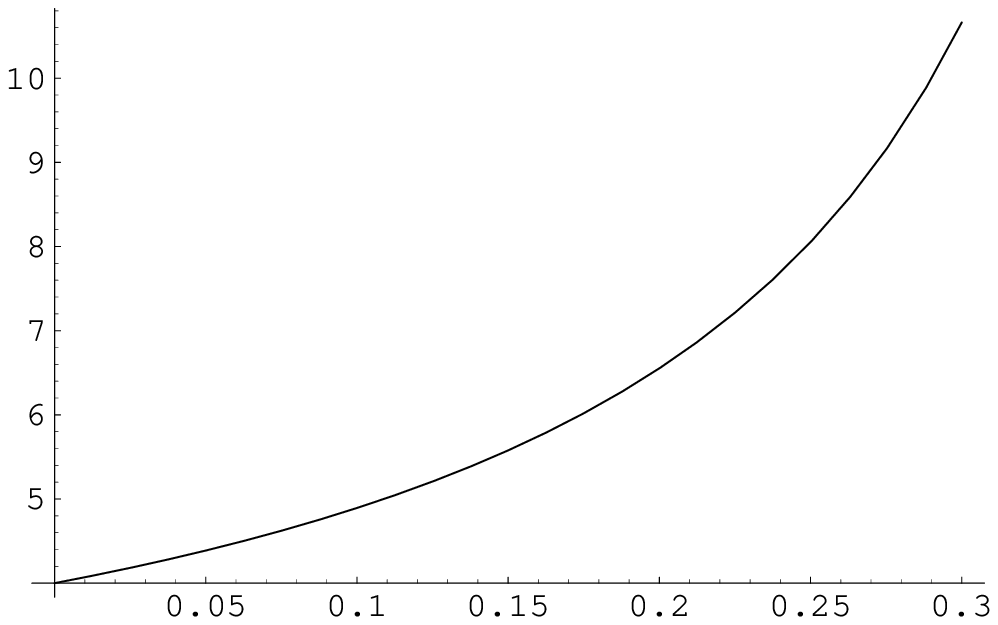}%
\caption{$A=4$. Graph of $\alpha(t)$.}
\label{f6}
}
\parbox{3in}{
\includegraphics[
height=2.0in,
width=2.8in
]%
{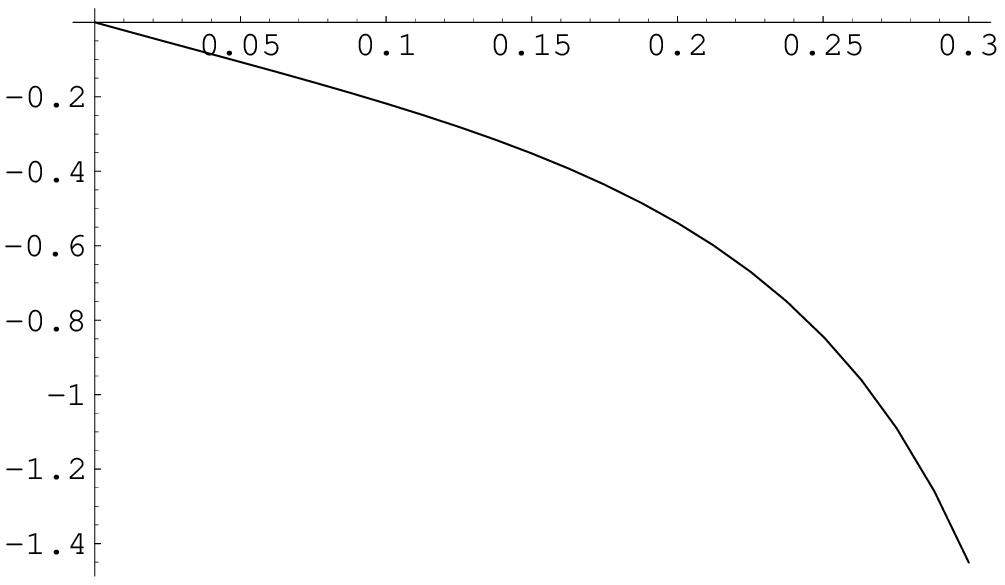}%
\caption{$A=4$. Graph of $\gamma(t)$.}
\label{f7}
}
\begin{center}
\includegraphics[
height=2.0in,
width=2.8in
]%
{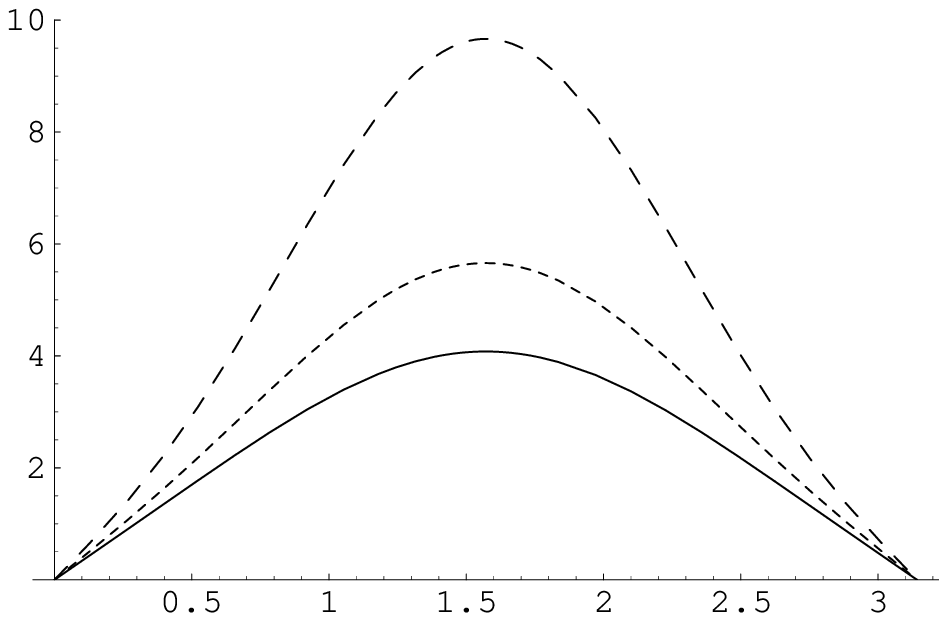}%
\caption{$A=4$. Graphs of $\fiap(t)(x)$ for $x \in (0, \pi)$ and $t=0.1$ (continuous line), $t=0.2$
(short dashes), $t=0.3$ (long dashes).}
\label{f8}
\end{center}
\parbox{3in}{
\includegraphics[
height=2.0in,
width=2.8in
]%
{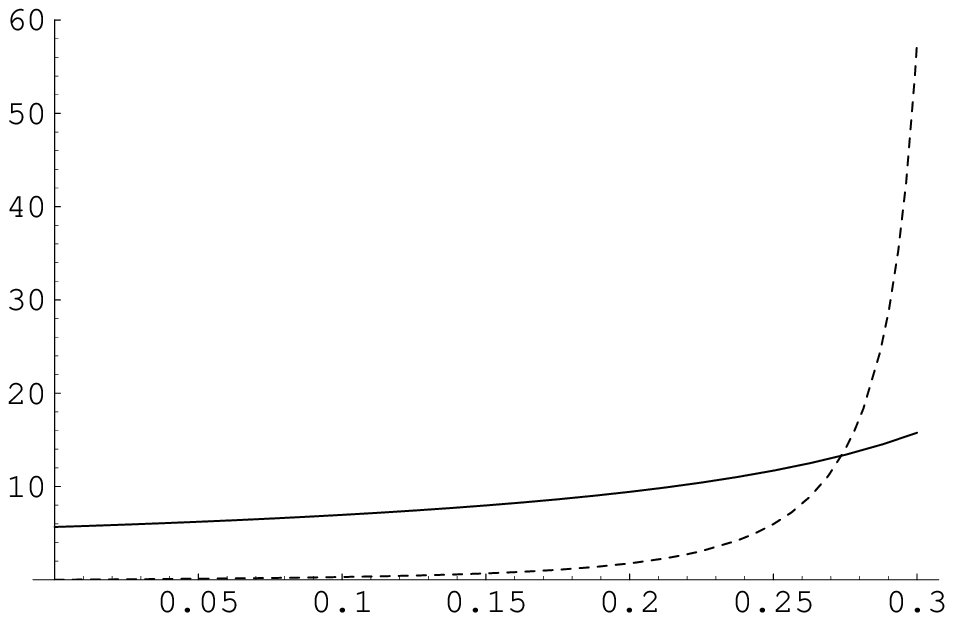}%
\caption{$A=4$. Graphs of $\| \fiap(t) \|$ (continuous line) and $R(t)$  (dashed line).}
\label{f9}
}
\parbox{3in}{
\includegraphics[
height=2.0in,
width=2.8in
]%
{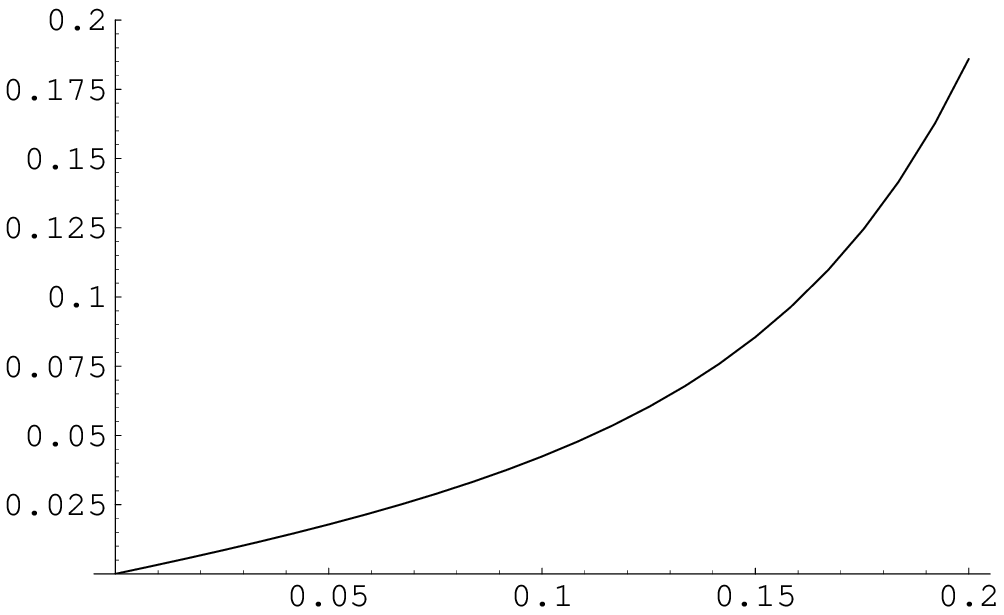}%
\caption{$A=4$. Graph of $R(t)/\| \fiap(t) \|$.}
\label{f10}
}
\end{figure}
\vfill \eject \noindent
\appendix
\section{Appendix. Sobolev spaces and pointwise product.}
\label{appe1}
We consider the space $H^1(\reali, \complessi) := \{ f : \reali \vain \complessi~|~f, f_x \in L^2(\reali) \}$
with the inner product $< f\,|\, g>  := < f \,|\, g>_{\! L^2} + < f_x \,|\, g_x >_{\! L^2}$ and
the corresponding norm $\|~\|$. This is
useful to treat the space $H^{1}_0(0,\pi)$ of Sect.\ref{appl} (made of \textsl{real} functions): in fact, there is
an $\reali$-linear, norm preserving inclusion
\beq H^{1}_0(0,\pi) \subset H^1(\reali, \complessi)~, \label{inc} \feq
where each $f \in H^1_0(0,\pi)$ is extended to the
full real axis setting $f(x) := 0$ for $x \not\in (0,\pi)$.
\parn
Both $H^1_0(0,\pi)$ and $H^1(\reali,\complessi)$ are closed
under the pointwise product, and $\| f g \| \leq$ const. $\| f \| \, \| g \|$
for all functions therein \cite{Run}. We claim the following:
\begin{prop}
\textbf{Proposition.} Consider the sharp (i.e., the minimum) constants $L, M$ in the inequalities
\beq \| f g \| \leq L \| f \| \, \| g \| \qquad \mbox{for all $f, g \in H^1_0(0,\pi)$}~; \label{applying} \feq
\beq \| f g \| \leq M \| f \| \, \| g \| \qquad \mbox{for all $f, g \in H^1(\reali, \complessi)$~.}
\label{special} \feq
Then
\beq 0.811 < L \leq M \leq 1~. \label{ellem} \feq
\end{prop}
\textbf{Proof.} i) $L \leq M$ follows readily
from \rref{inc}. \parn
ii) A lower bound for $L$ follows applying \rref{applying} with
$f = g = f_{\lambda}$, where
\beq f_{\la}(x) := e^{-\la | x - \pi/2 |} - e^{-\la \pi/2} \qquad (\lambda > 0)~,\feq
The norms $\| f_{\la} \|$, $\| f^2_{\la} \|$ are computed in an elementary way, and we get
a minorant of $L$ for each $\lambda$. The best lower bound is attained for
$\lambda$ close to $1.55$, and implies $L > 0.811$. \parn
iii) Let us prove that $M \leq 1$. To this purpose, we employ for
the complex functions $f$ on $\reali$ the Fourier transform
$(\FF f)(k) = {1 \over \sqrt{2 \pi}} \int_{\reali} d x~e^{- i k x} f(x)$; this
yields the representation
\beq H^1(\reali, \complessi) = \{ f \in L^2(\reali, \complessi)~|~+ \infty > \int_{\reali}
d k~(1 + k^2) | \FF f(k) |^2 = \| f \|^2~\}~, \feq
and sends pointwise product into $(1/\sqrt{2 \pi}) \times$ the convolution product $\ast$~.
Consider any two functions $f, g \in H^1(\reali, \complessi)$. Then the following holds:
\beq \| f g \|^2 = \int_{\reali} d k (1 + k^2) | \FF(f g)(k) |^2 = {1 \over 2 \pi}
\int_{\reali} d k (1 + k^2) | (\FF f \ast  \FF g)(k) |^2~; \label{laprec} \feq
\beq  (\FF f \ast  \FF g)(k)  = \int_{\reali} d h~ \FF f(k - h) \FF g (h)  = \feq
$$ = \int_{\reali} d h {1 \over \sqrt{1 + (k - h)^2} \sqrt{1 + h^2}}~
\left( \sqrt{1 + (k - h)^2}~\FF f(k - h) \sqrt{1 + h^2} \FF g (h) \right)~; $$
\beq {1 \over 2 \pi} | (\FF f \ast  \FF g)(k) |^2 \leq C(k) P(k)~, \label{dains} \feq
\beq C(k) := {1 \over 2 \pi} \int_{\reali} {d h \over (1 + (k - h)^2) (1 + h^2) } = {1 \over 4 + k^2}~, \feq
\beq P(k) := \int_{\reali} d h~ (1 + (k - h)^2) | \FF f (k - h) |^2
(1 + h^2) | \FF g(h) |^2~.\feq
Eq.\rref{dains} follows from H\"older's inequality $| \int d h~ U V |^2 \leq \left(\int d h | U |^2\right)
\left(\int d h~| V |^2 \right)$; inserting \rref{dains} into Eq.\rref{laprec}, we get
\beq \| f g \|^2 \leq \int_{\reali} \! \! \! d k (1 + k^2) C(k) P(k) \leq
\left(\sup_{k \in \reali} (1 + k^2) C(k) \right) \int_{\reali} \! \! \! d k~ P(k)~= \label{subs} \feq
$$ \hspace{6cm} = 1 \times \| f \|^2 \| g \|^2~. \hspace{6cm} \diamond $$
In \cite{MP} we have discussed the constants for more general inequalities related
to the pointwise product and to the spaces $H^n(\reali^d, \complessi)$. The upper bound $M \leq 1$ derived
now improves the result arising from \cite{MP} in the special case of the inequality \rref{special};
the method employed here to bind $M$ develops in a fully quantitative way an idea suggested in
\cite{Pos}.
\section{Appendix. Proof of Prop.\ref{kapl}.}
\label{appe2}
We keep all notations of Sect.\ref{appl}. The proof consists of the following steps: \parn
i) The function $Q$ of Eq.\rref{deq} can be seen as a continuous linear form,
both on $L^2(0,\pi)$ and on $\FFF$. For all $f \in \FFF$ and $t \in [0,+\infty)$, we easily infer
from \rref{semiheat} that
\beq Q(\UU(t) f) = e^{-t} Q(f)~. \label{lee} \feq
ii) Let $f \in L^{2 p}(0,\pi) (\subset L^2(0,\pi)$) and $f \geq 0$; then
\beq Q(f) \leq Q(f^p)^{1/p}~. \label{qfp} \feq
This follows taking $q$ such that $1/p + 1/q=1$, and
writing $Q(f) =\int_{0}^\pi d x~ u(x) v(x)$ with
$u(x) := ({1 \over 2} \sin x)^{1/q}$, $v(x) := ({1 \over 2} \sin x)^{1/p} f(x)$;
H\"older's inequality $\int u v \leq (\int u^q)^{1/q} (\int v^p)^{1/p}$ yields Eq.\rref{qfp}
({\footnote{Eq.\rref{qfp} is optimal, in this sense: the best constant in the inequality $Q(f) \leq C Q(f^p)^{1/p}$
for all nonnegative $L^{2 p}$ functions is $C=1$. This
is true even if we restrict the inequality to much smaller classes,
such as the nonnegative $C^{\infty}$, compactly supported  functions $f$ on $(0,\pi)$.}}). \parn
iii) We consider the maximal solution $\varphi : [0, \theta) \vain \FFF$ of
the Volterra problem \rref{eqvol} with datum $f_0 \in \FFF$, assuming $f_0 \geq 0$ and
$Q(f_0) > 1$; the nonnegativity of $f_0$ implies $\varphi(t) \geq 0$ for all $t$  (see, e.g., \cite{Caz}). We define
the (continuous) function
\beq t \in [0, \theta) \mapsto Q(t):= Q(\varphi(t))~,
\feq
and note that
\beq Q(t) \geq 0~, \qquad Q(t) \geq e^{- t} Q(f_0) + \int_{0}^t d s~ e^{-(t-s)} Q(s)^{p}~; \label{diffin} \feq
the first bound follows from $\varphi(t) \geq 0$, and the second one from
\rref{eqvol} \rref{lee} \rref{qfp}. \parn
iv) For each $n \in \naturali$, we define
a continuous function $S_n : [0, \vartheta) \vain \reali$ by
\beq S_0(t) := e^{-t} Q(f_0)~, \qquad S_{n+1}(t) := e^{-t} Q(f_0) + \int_{0}^t d s~e^{-(t-s)} S_n(s)^p~;
\label{th1} \feq
it is proved recursively that, for all $n \in \naturali$ and $t \in [0, \vartheta)$,
\beq Q(t) \geq S_n(t) \geq 0 \label{th2} \feq
(the first inequality depends on \rref{diffin}, the second one is elementary).
It is easily checked that the sequence of functions $(S_n)$ is a Cauchy sequence in the
topology of the uniform convergence on all compact subintervals $[0, \tau] \subset [0, \vartheta)$;
its $n \vain +\infty$ limit is a continuous function
$S : [0, \vartheta) \vain \reali$ such that, for all $t$ in this interval,
\beq S(t) = e^{- t} Q(f_0) + \int_{0}^t d s~ e^{-(t-s)} S(s)^{p}, \quad
Q(t) \geq S(t) \geq 0. \label{inteq} \feq
From the above integral equation, we see
that $S$ is in fact $C^1$, and fulfils the Cauchy problem
\beq {\dot S}(t) = S(t) \Big( S
(t)^{p-1} - 1 \Big)~, \qquad S(0)= Q(f_0)~; \label{todisc} \feq
this has a unique maximal solution (for nonnegative times),
denoted again with $t \mapsto S(t)$, which extends
the function considered up to now on $[0,\vartheta)$, and is given by
\beq \int_{Q(f_0)}^{S(t)} {d r \over r (r^{p-1} - 1)} = t \qquad \mbox{for $t \in [0,\TK)$,}
\qquad \TK := \int_{Q(f_0)}^{+\infty} {d r \over r (r^{p-1} - 1)}~; \feq
furthermore, $S(t) \vain +\infty$ for $t \vain \TK^{-}$. Computing the last integral,
we see that $\TK$ has the expression \rref{det1} in the statement of the theorem;
we know  that $\vartheta \leq \TK$, so the proof of Prop
\ref{kapl} is concluded. \fine
\vskip 0.2cm \noindent
\textbf{Acknowledgments.} We are grateful to the anonymous referees
for some suggestions that stimulated an improvement of the paper.
This work was partly supported by INdAM and by MIUR, COFIN 2001
Research Project "Geometry of Integrable Systems".

\end{document}